\documentclass[12pt,english]{article}
\usepackage[T1]{fontenc}
\usepackage[latin9]{inputenc}
\usepackage{geometry}
\usepackage{babel}
\usepackage{amsmath}
\usepackage{amssymb}
\usepackage{setspace}
\usepackage[unicode=true,pdfusetitle,
 bookmarks=true,bookmarksnumbered=false,bookmarksopen=false,
 breaklinks=false,pdfborder={0 0 1},backref=false,colorlinks=false]
 {hyperref}
\usepackage{breakurl}

\makeatletter
\numberwithin{equation}{section}
\newcommand{\lyxaddress}[1]{
\par {\raggedright #1
\vspace{1.4em}
\noindent\par}
}

\usepackage{cite}

\makeatother

\begin{document}

\title{BPS Solutions of Six-Dimensional (1,0) Supergravity Coupled to Tensor
Multiplets}

\author{Huibert het Lam and Stefan Vandoren}

\maketitle
\vspace{-1cm}
\begin{singlespace}
\lyxaddress{\begin{center}
\textit{Institute for Theoretical Physics and Center for Extreme
Matter and Emergent Phenomena,} \\ \smallskip
\textit{Utrecht University, 3508 TD Utrecht,
The Netherlands}
\par
\bigskip
\medskip
\href{mailto:H.hetLam@uu.nl}{H.hetLam@uu.nl, }
\href{mailto:S.J.G.Vandoren@uu.nl}{S.J.G.Vandoren@uu.nl}
\end{center}}
\end{singlespace}

\vspace{2cm}

\begin{abstract}
We derive a general local form for supersymmetric solutions of  six-dimensional
$(1,0)$ supergravity coupled to an arbitrary number of tensor multiplets.
%These solutions arise naturally from F-theory compactifications on
%elliptically fibered Calabi-Yau threefolds. 
We consider some special
cases in which the resulting equations can be solved explicitly. In
particular we derive black string solutions and calculate their entropy.
Upon reducing to five dimensions they yield spinning black hole solutions.
We also discuss BPS pp-waves and black string solutions with traveling waves.
Lastly, as an application, we study the attractor mechanism in this
theory. 
\end{abstract}
\thispagestyle{empty}

\newpage{}

\tableofcontents{}

\newpage{}

\section{Introduction}

The phase space of (quantum) gravity solutions in dimensions larger than four is intricate and has a rich structure, see e.g. \cite{Emparan:2008eg,Kunduri:2013ana} for some reviews on black holes and horizons in various dimensions. Whereas in asymptotically flat four spacetime dimensions, horizon topologies are unique to be $S^2$, in five dimensions one can have black holes and black rings \cite{Emparan:2001wn}, with horizons $S^3$ and $S^1 \times S^2$ respectively. Also quotient topologies exist, such as Lens space horizons \cite{Kunduri:2014kja,Kunduri:2016xbo,Tomizawa:2016kjh}. These objects can be made BPS in supergravity and are embeddable in string theory where one can provide a microscopic description. In six dimensions, near horizon geometries have been classified in (1,0) supergravity coupled to tensor multiplets and hypermultiplets (and no vector multiplets) \cite{Gutowski:2003rg,Akyol:2011mh}, and there are more possibilities. The focus of this paper, the case where the hypermultiplets are frozen, only yields near horizon geometries locally given by either $\mathbb{R}^{1,1}\times T^{4},$ $\mathbb{R}^{1,1}\times K3$
or $AdS_{3}\times S^{3}.$  Some of the objects with a horizon easily follow from uplifts from 5d to 6d. The most well studied case is of course the 6d BPS black string with horizon $S^1 \times S^3$ and near horizon geometry $AdS_3\times S^3$, arising from the uplift of a 5d spherical black hole. The microscopic description in string theory was first given in \cite{Strominger:1996sh,Breckenridge:1996is}. One can also uplift a 5d black ring and we will see that it also has horizon $S^1\times S^3$ and near horizon geometry $AdS_{3} \times S^3$. Besides black objects with horizons, there are also smooth horizonless BPS (microstate-) geometries (see \cite{Bena:2011dd,Niehoff:2013kia,Giusto:2013rxa,Bena:2015bea,deLange:2015gca,Bena:2017xbt} for an incomplete list of references), and an interesting class of 6d BPS pp-wave solutions that we will study in this paper. Many other solutions can be found by superposing waves on black strings, and one can play with various kinds of asymptotics. 

So far general BPS solutions of $(1,0)$ supergravity have mainly been studied in the minimal case
\cite{Gutowski:2003rg} and in the case with the gravity multiplet coupled to one or two tensor multiplets (gauged and ungauged), see e.g.
\cite{Cariglia:2004kk,Martelli:2004xq,Bena:2011dd,Bobev:2012af,Niehoff:2012wu,Niehoff:2013kia,Giusto:2013rxa,Bena:2015bea,deLange:2015gca,Bena:2017xbt}. In this paper we generalize this to $n_T$ tensor multiplets, for any $n_T$. One of the new ingredients is the scalars in the tensor multiplets which can have nontrivial profiles and that are subject to a 6d attractor mechanism, as we will see. 

Six-dimensional $(1,0)$ supergravity coupled to matter multiplets arises from the compactification
of F-theory on elliptically fibered Calabi-Yau threefolds or from
truncations of type IIB on $T^{4}$ or $K_{3}.$ Of particular interest
are BPS black string solutions of this theory, as they yield five-dimensional
black holes upon further compactification on a circle. The F-theory microscopics
have been studied in \cite{Vafa:1997gr,Haghighat:2015ega,Lawrie:2016axq}. The near
horizon geometry leads to new two-dimensional $(0,4)$ CFTs that have
been recently investigated in \cite{Haghighat:2015ega,Couzens:2017way,Couzens:2017nnr}. Compactifying F-theory on an elliptically fibered non-singular
Calabi-Yau threefold results in a gravity multiplet, $n_{T}=h^{1,1}(B_{2})-1$
tensor multiplets and $n_{H}=h^{2,1}(CY_{3})+1$
hypermultiplets, where $h^{1,1}(B_{2})$ and $h^{2,1}(CY_{3})$ are
hodge numbers of the base and threefold respectively \cite{Vafa:1996xn,Morrison:1996na,Morrison:1996pp}. For generic elliptic fibrations, one has $n_V=h^{1,1}(CY_3)-h^{1,1}(B_{2})-1$ vector multiplets and an anomaly cancellation condition. Here we truncate all vectors and set the corresponding charges to zero. 
One approach to study F-theory is via its connection with M-theory:
one gets the effective 6d $(1,0)$ theory by reducing M-theory on the
Calabi-Yau threefold to five dimensions and then lifting it up to
six dimensions \cite{Bonetti:2011mw}. To do this, one has to take into account
one-loop contributions coming from the reduction on the circle \cite{Bonetti:2013ela}. So in a way six-dimensional solutions are proper F-theory solutions and it might be interesting
to see what we can learn from studying the connection between them
and their five-dimensional counterparts. 

More technically, our analysis starts with deriving a general local form for supersymmetric solutions where we use methods that have been applied before to four-dimensional theories
\cite{Gibbons:1982fy,Tod:1983pm,Tod:1995jf,Denef:2000nb,Meessen:2006tu,Huebscher:2006mr}, five-dimensional theories \cite{Gauntlett:2002nw,Gauntlett:2003fk,Gutowski:2004yv,Bellorin:2006yr}
and minimal six-dimensional supergravity \cite{Gutowski:2003rg}.
The strategy is always the same. Starting from a Killing spinor $\epsilon$
one constructs bosonic objects quadratic in $\epsilon,$ the so-called
bilinears, such as the vector $X^{\mu}=\bar{\epsilon}\gamma^{\mu}\epsilon$.
These bilinears have certain properties since $\epsilon$ is a Killing
spinor. The vector $X^{\mu}$ for instance turns out to be a Killing
vector in all cases. Using these bilinears, the local form of the
solutions can be identified and this can be used to simplify the equations
of motion. It turns out that the resulting equations in minimal $N=2,$
$D=4$ supergravity can actually be solved completely \cite{Tod:1983pm}.
In five-dimensional minimal supergravity the solutions fall into two
classes \cite{Gauntlett:2002nw}. In the first class, the vector $X$
is null and the solutions are plane-fronted waves expressed in harmonic
functions on $\mathbb{R}^{3}$. In the second class, the vector $X$
is timelike and the equations of motion can not be solved completely,
but the equations are simplified significantly such that one only
has to make an ansatz for the remaining variables. Solutions of  supergravity
coupled to an arbitrary number of vector multiplets have similar properties
as in the minimal case \cite{Gutowski:2004yv}. In six-dimensional
minimal supergravity the Killing vector $X$ is always null \cite{Gutowski:2003rg}
and as one can expect, based on the five-dimensional analysis, the
equations of motion can not be solved completely in the most general
case. 

The bosonic field content in the six-dimensional theory consists of the metric, a two-form with anti-self-dual field strength in the gravity multiplet, and $n_{T}$ two-forms with self-dual field strength and $n_{T}$ scalars in the tensor multiplets.
The constraints that supersymmetry puts on the field content have
already been derived in \cite{Akyol:2010iz}, so like in \cite{Gutowski:2003rg}
we will introduce coordinates and reduce the equations of motion using
these local expressions. Using coordinates $(u,v,x^{m}),$ $m=1,...,4$ we find that solutions
are not dependent on $v$ and can be expressed in terms of a $u-$dependent
base manifold $\mathcal{B}$. In general the base space $\mathcal{B}$
exhibits a non-integrable hyper-Kähler  structure. Using the form
of the solutions that one gets from requiring one Killing spinor,
all the equations can be expressed in terms of bosonic quantities
on $\mathcal{B}.$ These equations are not easy to solve in full generality,
but it is still easier to find solutions by substituting an ansatz
in these equations than to start with an ansatz for the complete field
content. 

The resulting equations will be studied in two cases where the base
space becomes hyper-Kähler. One of those cases arises when the solution
is $u-$independent. In that case we also take $\mathcal{B}$ to be
Gibbons-Hawking \cite{Gibbons:1979zt} and it turns out that the solution
is completely determined in terms of $6+2n_{T}$ harmonic functions
on $\mathbb{R}^{3}$. This is one of the main new results of our analysis. When one takes these solutions to be multi-centered
Gibbons-Hawking, requiring the absence of Dirac string-like singularities
gives restrictions on the relative positions of the centers. Just
as in the minimal case \cite{Crichigno:2016lac} there is a symplectic
group that sends solutions to solutions preserving regularity, but here the $Sp(6)$ gets enlarged to $Sp(6+2n_T)$. We
also construct the macroscopic black string solutions in F-theory backgrounds $\mathbb{R}\times S^{1}\times\mathbb{R}^{4}\times CY_{3}$
with a D3-brane wrapped on $S^{1}\times C,$ where $C$ is a curve
in the base of the Calabi-Yau threefold. 

We finish by studying the attractor mechanism \cite{Ferrara:1995ih,Ferrara:1996dd}
in this theory. This has partly been done in \cite{Andrianopoli:2007kz,Ferrara:2008xz}
but only the near horizon analysis. Here we derive a 
flow equation on the tensor branch for $u-$independent solutions. We look at simplifying assumptions needed to understand the
attractor values as the optimization of the central charge, and we apply this to the one-centered Gibbons-Hawking class of solutions that include BPS black strings. Some of the latter solutions have also been treated in \cite{deAntonioMartin:2012bi}.

This paper is organized as follows. In section \ref{sec:Setting} we
describe the field content of (1,0) supergravity with a gravity
multiplet and tensor multiplets, we list the conditions
the existence of a Killing spinor puts on the field content, and we
introduce the equations of motion that are not implied by integrability
conditions. In section \ref{sec:Supersymmetric-Solutions} we introduce
coordinates and reduce the equations of motion to equations in terms
of bosonic quantities on $\mathcal{B}.$ Section \ref{sec:Special-Cases}
then solves the resulting equations under certain assumptions. Here
we also discuss how the theory reduces to five dimensions and look at black strings and other objects with a horizon. We finish this section by looking at examples of pp-waves. Section \ref{sec:Attractor-Mechanism} describes the attractor mechanism for this theory; here we also derive
a flow equation for $u-$independent solutions. In section \ref{sec:Summary-and-Discussion}
we then summarize and give some suggestions for future work.

\paragraph{Notation.}

Since we use a lot of different notation in this paper, we give a
short overview:
\begin{itemize}
\item $M,N=1,...,n_{T}$ where $n_{T}$ is the number of tensor multiplets,
\item $\alpha,$ $\beta,...=1,...,n_{T}+1$ denote field content in six
dimensions,
\item $\mu,\nu,...=0,...,5$ denote coordinates or the vielbein of the six-dimensional
metric,
\item $i,j,k,l=1,...,4$ first denote part of the vielbein of the six-dimensional
space and from section \ref{sec:Supersymmetric-Solutions} onwards
the vielbein of the base manifold $\mathcal{B},$
\item $m,n,...=1,...,4$ denote the coordinates of the base manifold $\mathcal{B},$ 
\item $a,b,c,d=1,...,3$ denote either the three two-forms in the almost
hyper-Kähler structure, or part of the vielbein of a Gibbons-Hawking
metric,
\item $I,J,...=0,...,n_{T}+1$ denote field content in five dimensions,
\item objects with a hat $\hat{\cdot}$ live in six dimensions,
\item objects with a tilde $\tilde{\cdot}$ live in the four-dimensional
base space,
\item $\star_{6}$ is the hodge star in six dimensions,
\item $\star_{4}$ is the hodge star in the four-dimensional base space.
\end{itemize}

\paragraph{Note added.} During the submission process of this paper, we learned of similar and independent work \cite{Cano:2018wnq} that has some overlap with ours.

\section{Setting \label{sec:Setting}}

We consider six-dimensional (1,0) supergravity coupled to $n_{T}$ tensor multiplets  \cite{Romans:1986er,Sagnotti:1992qw,Ferrara:1997gh}. Vector and hypermultiplets
can be added to ensure an F-theory embedding, but we will set them
to zero in the solutions we consider in this paper. The idea is to
study the BPS structure of the tensor branch. The bosonic content
of the gravity multiplet consists of a graviton and a two-form with anti-self-dual field strength. Every tensor multiplet contains a two-form with self-dual field strength and
a scalar. We will denote the six-dimensional two-forms by $\hat{B}^{\alpha},$
where $\alpha=1,...,n_{T}+1.$ The scalars of the tensor multiplets
parametrize the coset space $SO(1,n_{T})/SO(n_{T}).$ A convenient
way to describe them is by an $SO(1,n_{T})$ matrix 
\begin{equation}
S=\left(\begin{array}{c}
j_{\alpha}\\
x_{\alpha}^{M}
\end{array}\right),\ M=1,...,n_{T},
\end{equation}
whose matrix elements satisfy the constraints 
\begin{eqnarray}
j_{\alpha}j_{\beta}-\sum_{M}x_{\alpha}^{M}x_{\beta}^{M} & = & \Omega_{\alpha\beta},\nonumber \\
j_{\alpha}j^{\alpha} & = & 1,\\
x_{\alpha}^{M}j^{\alpha} & = & 0,\nonumber 
\end{eqnarray}
where $\Omega_{\alpha\beta}=\mathrm{diag}(1,-1,...,-1)$ is used to
lower and raise $\alpha$ indices. The field strengths corresponding
to the two-forms are given by $\hat{G}^{\alpha}=d\hat{B}^{\alpha}$
and their relation to the anti-self-dual tensor $H$ of the gravity
multiplet and the self-dual tensors $H^{M}$ of the tensor multiplets
is given by
\begin{equation}
\hat{G}^{\alpha}=j^{\alpha}H-\Omega^{\alpha\beta}\sum_{M}x_{\beta}^{M}H^{M}.\label{relation three forms G and H}
\end{equation}
In terms of the three-forms $\hat{G}^{\alpha}$ the self-duality condition
can be written as 
\begin{equation}
g_{\alpha\beta}\star_{6}\hat{G}^{\beta}=-\Omega_{\alpha\beta}\hat{G}^{\beta},
\end{equation}
where
\begin{equation}
g_{\alpha\beta}=2j_{\alpha}j_{\beta}-\Omega_{\alpha\beta}
\end{equation}
is a positive definite metric. 

The self-duality condition on the three-forms makes it hard to construct a covariant action functional from which all equations of motion follow. These actions do exist \cite{Pasti:1996vs,DallAgata:1997yxl}, but one has to introduce auxiliary fields. Another approach is a pseudo-action
from which equations of motion follow that then still have to be supplemented by the self-duality
condition. The bosonic part of
the pseudo-action that is relevant for this paper is given by \cite{Ferrara:1997gh}
\begin{equation}
S=\int_{\mathcal{M}_{6}}\frac{1}{2}R\star_{6}1-\frac{1}{4}g_{\alpha\beta}\hat{G}^{\alpha}\wedge\star_{6}\hat{G}^{\beta}+\frac{1}{2}\Omega_{\alpha\beta}dj^{\alpha}\wedge\star_{6}dj^{\beta}.
\end{equation}
The equations of motion are then equal to
\begin{eqnarray}
R_{\mu\nu} & = & \frac{1}{4}g_{\alpha\beta}\hat{G}_{\mu}^{\alpha\ \rho\lambda}\hat{G}_{\nu\rho\lambda}^{\beta}-\frac{1}{24}\hat{g}_{\mu\nu}g_{\alpha\beta}\hat{G}_{\rho\lambda\sigma}^{\alpha}\hat{G}^{\beta\ \rho\lambda\sigma}-\Omega_{\alpha\beta}\partial_{\mu}j^{\alpha}\partial_{\nu}j^{\beta},\nonumber \\
x_{\alpha}^{M}d\left(\star_{6}dj^{\alpha}\right) & = & -x_{\alpha}^{M}j_{\beta}\hat{G}^{\alpha}\wedge\star_{6}\hat{G}^{\beta},\nonumber \\
d\left(g_{\alpha\beta}\star_{6}\hat{G}^{\beta}\right) & = & 0,\\
g_{\alpha\beta}\star_{6}\hat{G}^{\beta} & = & -\Omega_{\alpha\beta}\hat{G}^{\beta},\nonumber \\
j_{\alpha}j^{\alpha} & = & 1.\nonumber 
\end{eqnarray}
Using the self-duality condition, this set of equations is equivalent
to
\begin{eqnarray}
R_{\mu\nu} & = & \frac{1}{4}g_{\alpha\beta}\hat{G}_{\mu}^{\alpha\ \rho\lambda}\hat{G}_{\nu\rho\lambda}^{\beta}-\Omega_{\alpha\beta}\partial_{\mu}j^{\alpha}\partial_{\nu}j^{\beta},\nonumber \\
x_{\alpha}^{M}d\left(\star_{6}dj^{\alpha}\right) & = & x_{\alpha}^{M}j_{\beta}\hat{G}^{\alpha}\wedge\hat{G}^{\beta},\nonumber \\
d\hat{G}^{\alpha} & = & 0,\label{Equations of Motion before susy}\\
g_{\alpha\beta}\star_{6}\hat{G}^{\beta} & = & -\Omega_{\alpha\beta}\hat{G}^{\beta},\nonumber \\
j_{\alpha}j^{\alpha} & = & 1.\nonumber 
\end{eqnarray}
Note that the equation of motion for the three-forms becomes the Bianchi identity.

\subsection{$N=1$ restrictions\label{subsec:-restrictions}}

The BPS equations for six-dimensional (1,0) supergravity with a gravity
multiplet and tensor multiplets are \cite{Akyol:2010iz}
\begin{eqnarray}
\left(\nabla_{\mu}-\frac{1}{8}H_{\mu\nu\rho}\gamma^{\nu\rho}\right)\epsilon & = & 0,\nonumber \\
\left(\frac{i}{2}T_{\mu}^{M}\gamma^{\mu}-\frac{i}{24}H_{\mu\nu\rho}^{M}\gamma^{\mu\nu\rho}\right)\epsilon & = & 0,\label{eq:tensorini bps equation}
\end{eqnarray}
where 
\begin{eqnarray}
T_{\mu}^{M} & \equiv & x_{\alpha}^{M}\partial_{\mu}j^{\alpha}.
\end{eqnarray}
We will consider geometries with $N=1$ supersymmetry. In \cite{Akyol:2010iz}
the restrictions that the Killing spinor puts on the geometry have
been worked out which we will discuss in this section. 

On our geometry we introduce a vielbein $\hat{e}^{0},$ ..., $\hat{e}^{5}$,
the null-forms\footnote{We relabeled the vielbein of \cite{Akyol:2010iz} to make notation
easier later on.}
\begin{eqnarray}
e^{-} & = & \frac{1}{\sqrt{2}}\left(-\hat{e}^{0}+\hat{e}^{5}\right),\nonumber \\
e^{+} & = & \frac{1}{\sqrt{2}}\left(\hat{e}^{0}+\hat{e}^{5}\right)
\end{eqnarray}
and choose the orientation $\epsilon^{-+1234}=1.$ Using the null-vielbein
$e^{-}$, $e^{+},$ $\hat{e}^{i}$, where $i=1,...,4,$ we can write
the metric as
\begin{equation}
ds_{6}^{2}=2e^{-}e^{+}+\delta_{ij}\hat{e}^{i}\hat{e}^{j}.\label{metric susy}
\end{equation}
We choose the orientation of the directions perpendicular to the light-cone
direction $\epsilon^{ijkl}=\epsilon^{-+ijkl}$ (note that \cite{Akyol:2010iz}
uses the opposite orientation). 

The Killing spinor can then be used to construct the bilinears of
the geometry, which turn out to be given by $e^{-}$ and
\begin{equation}
e^{-}\wedge I^{1},\ \ e^{-}\wedge I^{2},\ \ e^{-}\wedge I^{3},
\end{equation}
where the two-forms $I^{a}$ for $a\in\{1,2,3\}$ take the form
\begin{eqnarray}
I^{1} & = & -\left(\hat{e}^{1}\wedge\hat{e}^{3}+\hat{e}^{2}\wedge\hat{e}^{4}\right),\nonumber \\
I^{2} & = & -\left(\hat{e}^{1}\wedge\hat{e}^{2}-\hat{e}^{3}\wedge\hat{e}^{4}\right),\label{eq:hyper kahler structure}\\
I^{3} & = & -\left(\hat{e}^{1}\wedge\hat{e}^{4}-\hat{e}^{2}\wedge\hat{e}^{3}\right).\nonumber 
\end{eqnarray}
These two-forms are anti-self-dual on the directions perpendicular
to the light-cone direction and satisfy the algebra of the imaginary unit quaternions on a four-manifold with metric $\delta_{ij} \hat{e}^{i} \hat{e}^{j}$:
\begin{equation}
\left(I^{a}\right)_{\ k}^{i}\left(I^{b}\right)_{\ j}^{k}=-\delta^{ab}\delta_{\ j}^{i}+\epsilon_{\ \ \,c}^{ab}\left(I^{c}\right)_{\ j}^{i}.
\end{equation} 
Note that the vierbein chosen in (\ref{eq:hyper kahler structure}) is special and that the forms $I^{a}$ might look different when another vierbein is used. 

The conditions that the gravitino Killing
spinor equation imposes on the spacetime geometry can be rewritten
as
\begin{eqnarray}
\nabla_{\mu}e_{\nu}^{-} & = & \frac{1}{2}H_{\ \ \mu\nu}^{\lambda}e_{\lambda}^{-},\label{Gravitino condtion 1}\\
\nabla_{\mu}\left(e^{-}\wedge I^{a}\right)_{\nu\lambda\rho} & = & \frac{1}{2}H_{\ \ \mu\nu}^{\sigma}\left(e^{-}\wedge I^{a}\right)_{\sigma\lambda\rho}+\frac{1}{2}H_{\ \ \mu\lambda}^{\sigma}\left(e^{-}\wedge I^{a}\right)_{\nu\sigma\rho}+\frac{1}{2}H_{\ \ \mu\rho}^{\sigma}\left(e^{-}\wedge I^{a}\right)_{\nu\lambda\sigma}.\nonumber \\
\label{gravitino condition 2}
\end{eqnarray}
Condition (\ref{Gravitino condtion 1}) implies that the vector $X$
dual to $e^{-}$ is a Killing vector. Furthermore, condition (\ref{Gravitino condtion 1}),
the $\mu=-$ component of condition (\ref{gravitino condition 2})
and the anti-self-duality of $H$ imply that 
\begin{equation}
H=e^{+}\wedge de^{-}-\frac{1}{16}\left(I_{kl}^{a}\nabla_{-}I^{b\ kl}\right)\epsilon_{ab}^{\ \ \,c}I_{c\ ij}e^{-}\wedge\hat{e}^{i}\wedge\hat{e}^{j}+\frac{1}{6}\left(de^{-}\right)_{-l}\epsilon_{\ ijk}^{l}\hat{e}^{i}\wedge\hat{e}^{j}\wedge\hat{e}^{k}.\label{eq:Tensor gravity multiplet}
\end{equation}
The $\mu=+$ component of condition (\ref{gravitino condition 2})
implies that
\begin{equation}
\nabla_{+}I^{a}=0\label{+ cov derivative of two-forms}\ ,
\end{equation}
and the $\mu=i$ components give further restrictions on the covariant
derivatives of $I^{a}$. 
The tensorini Killing spinor equation implies that the scalars of
the tensor multiplets are invariant under the isometry $X$ \cite{Akyol:2010iz}.
Combining this equation with the self-duality condition yields that
the three-forms of the tensor multiplets take the following form \cite{Akyol:2010iz}:

\begin{eqnarray}
H^{M} & = & \frac{1}{2}H_{-ij}^{M}e^{-}\wedge\hat{e}^{i}\wedge\hat{e}^{j}+T_{i}^{M}e^{-}\wedge e^{+}\wedge\hat{e}^{i}+\frac{1}{6}T_{l}^{M}\epsilon_{\ ijk}^{l}\hat{e}^{i}\wedge\hat{e}^{j}\wedge\hat{e}^{k},\label{Tensors tensor multiplet}
\end{eqnarray}
where $\frac{1}{2}H_{-jk}^{M}e^{-}\wedge\hat{e}^{j}\wedge\hat{e}^{k}$
are self-dual tensors on the directions transverse to the light cone
part.

Notice that

\begin{equation}
\sum_{M}x_{\beta}^{M}T_{i}^{M}=-\Omega_{\alpha\beta}\partial_{i}j^{\alpha}.
\end{equation}
From (\ref{relation three forms G and H}), (\ref{eq:Tensor gravity multiplet})
and (\ref{Tensors tensor multiplet}) we then find that
\begin{eqnarray}
\hat{G}^{\alpha} & = & j^{\alpha}e^{+}\wedge de^{-}+e^{-}\wedge e^{+}\wedge dj^{\alpha}+\frac{1}{6}\left[j^{\alpha}\left(de^{-}\right)_{-l}+\left(dj^{\alpha}\right)_{l}\right]\epsilon_{\ ijk}^{l}\hat{e}^{i}\wedge\hat{e}^{j}\wedge\hat{e}^{k}\nonumber \\
 &  & -\left[\frac{1}{8}j^{\alpha}\left(I_{kl}^{a}\nabla_{-}I^{b\ kl}\right)\epsilon_{ab}^{\ \ \,c}\right]e^{-}\wedge I_{c}+e^{-}\wedge H_{\mathrm{SD}}^{\alpha}\ ,\label{threeforms}
\end{eqnarray}
where 
\begin{equation}
H_{\mathrm{SD}}^{\alpha}\equiv-\frac{1}{2}\Omega^{\alpha\beta}\sum_{M}x_{\beta}^{M}H_{-ij}^{M}\hat{e}^{i}\wedge\hat{e}^{j}
\end{equation}
is a self-dual two-form on the directions transverse to the light cone part.
Notice that $j_{\alpha}H_{\mathrm{SD}}^{\alpha}=0$. Using the Bianchi identity we calculate
\begin{eqnarray}
\mathcal{L}_{X}\hat{G}^{\alpha} & = & i_{X}d\hat{G}^{\alpha}+di_{X}\hat{G}^{\alpha}=d\left(j^{\alpha}de^{-}+dj^{\alpha}\wedge e^{-}\right)=0\ .
\end{eqnarray}
Condition (\ref{gravitino condition 2}) combined with the anti-self-duality
of $H$ and $I^{a}$ imply that $d\left(e^{-}\wedge I^{a}\right)=0$
such that also $\mathcal{L}_{X}\left(e^{-}\wedge I^{a}\right)=0\ .$
Hence $X$ generates a symmetry of the full solution.

\paragraph*{Summary.}

Using the vielbein $e^{-},$ $e^{+},$ $\hat{e}^{i}$ the field content
takes the form (\ref{metric susy}) and (\ref{threeforms}). Furthermore,
we have the two-forms (\ref{eq:hyper kahler structure}) that satisfy
the condition (\ref{+ cov derivative of two-forms}) and the $\mu=i$
components of (\ref{gravitino condition 2}). The null-vector $X$
dual to $e^{-}$ is a symmetry of the full solution. 

\subsection{Equations of motion}

From the integrability conditions it follows that if the BPS equations
(\ref{eq:tensorini bps equation}) and the Bianchi identity are satisfied,
the scalar equations of motion and all but the $--$ component of
the Einstein equation are obeyed (see appendix \ref{sec:Integrability-conditions}).
Hence the set of equations of motion (\ref{Equations of Motion before susy})
reduces to
\begin{eqnarray}
R_{--} & = & \frac{1}{4}g_{\alpha\beta}\hat{G}_{-}^{\alpha\ \rho\lambda}\hat{G}_{-\rho\lambda}^{\beta}-\Omega_{\alpha\beta}\partial_{-}j^{\alpha}\partial_{-}j^{\beta}\ ,\label{Einstein Equation}\\
d\hat{G}^{\alpha} & = & 0\ ,\label{Bianchi identity}\\
g_{\alpha\beta}\star_{6}\hat{G}^{\beta} & = & -\Omega_{\alpha\beta}\hat{G}^{\beta}\ ,\label{Self-duality condition}\\
j_{\alpha}j^{\alpha} & = & 1\ .\label{Restriction on the scalars}
\end{eqnarray}
Note that in the solution of the BPS equations (\ref{threeforms}),
the self-duality condition has already been taken care of. 

\section{Supersymmetric solutions \label{sec:Supersymmetric-Solutions}}

In this section we first use the existence of a null Killing
vector to introduce coordinates on the geometry. After this, we
rewrite the Bianchi identity and the Einstein equation using these
coordinates.

\subsection{Introduction of coordinates \label{subsec:Introduction-of-coordinates}} 

The existence of a null Killing vector $X$ can be used to introduce
coordinates on our spacetime. We implement this here following similar
steps as in \cite{Gutowski:2003rg}. First a hypersurface $\mathcal{S}$
has to be picked that is nowhere tangent to $X$. One then has to
pick a vector $Y$ that satisfies 
\begin{equation}
\hat{g}(Y,X)=1\ ,\ \ \ \hat{g}(Y,Y)=0\ ,\label{definition Y}
\end{equation}
on $\mathcal{S}.$ This vector $Y$ needs to be propagated off $\mathcal{S}$
by solving $\mathcal{L}_{X}Y=0.$ The properties (\ref{definition Y})
still hold since $X$ is Killing. The vectors $X$ and $Y$ commute,
hence they must be tangent to a four-parameter family of two-dimensional
surfaces $\Sigma_{2}(x^{m}),$ where $m=1,...,4.$ The vector $X$
is a null Killing vector, so should be tangent to affinely parametrized
null geodesics. Define the coordinate $v$ to be this affine parameter
along the geodesics and choose another coordinate $u$ such that $u,v$
are coordinates on the surfaces $\Sigma_{2}.$ We can then write 
\begin{eqnarray}\label{XandY}
X & = & \partial_{v}\ ,\nonumber \\
Y & = & H\left(\partial_{u}-\frac{1}{2}\mathcal{F}\partial_{v}\right)\ ,
\end{eqnarray}
for functions\footnote{This notation might be a bit confusing since we already used $H$
for the three-form in the gravity multiplet, but to more easily compare
with other literature, we will keep it. } $H$ and $\mathcal{F}$ independent of $v$. We will assume that (locally) $H>0$ since we can send $u \rightarrow -u$ if necessary. Using the properties of $X$ and $Y$ it can be shown \cite{Gutowski:2003rg} that the metric takes the form
\begin{equation}
ds_{6}^{2}=2H^{-1}\left(du+\beta\right)\left(dv+\omega+\frac{1}{2}\mathcal{F}\left(du+\beta\right)\right)+Hds_{4}^{2}\ ,\label{coodinates metric}
\end{equation}
where $ds_{4}^{2}=h_{mn}dx^{m}dx^{n}$ is the metric of a base space
$\mathcal{B}$ and $\omega,$ $\beta$ are one-forms on $\mathcal{B}$
\cite{Gutowski:2003rg}. The functions $H$ and $\mathcal{F},$ the
one-forms $\omega$ and $\beta$ and the metric $h_{mn}$ only depend
on $u$ and $x^{m}$ (since $X$ is a Killing vector). The one-forms
$e^{-},$ $e^{+}$ in (\ref{metric susy}) take the form
\begin{eqnarray}
e^{-} & = & H^{-1}\left(du+\beta\right),\nonumber \\
e^{+} & = & dv+\omega+\frac{1}{2}\mathcal{F}He^{-}.\label{eq:eplus and eminus}
\end{eqnarray}
Notice also that the relation between $\hat{e}^{i}$ and a vierbein
$\tilde{e}^{i}$ of $\mathcal{B}$ is given by $\hat{e}^{i}=H^{1/2}\tilde{e}^{i}$
which implies $\partial_{\hat{i}}=H^{-1/2}\partial_{i}$. From now
on, the labels $i,j,k,l$ will refer to the vielbein on the base space. 
\begin{flushleft}
Let us define anti-self-dual forms on $\mathcal{B}$ by 
\begin{equation}
J^{a}=H^{-1}I^{a}.
\end{equation}
These satisfy the algebra
\begin{equation}
\left(J^{a}\right)_{\ \ p}^{m}\left(J^{b}\right)_{\ \,n}^{p}=\epsilon_{\ \ \,c}^{ab}\left(J^{c}\right)_{\ \ n}^{m}-\delta^{ab}\delta_{n}^{m},\label{algebra kahler structure}
\end{equation}
where the indices $m,n,...$ have been raised with $h^{mn}.$ Hence,
these two-forms yield an almost hyper-Kähler structure on $\mathcal{B}.$
\par\end{flushleft}

Following \cite{Gutowski:2003rg} we  introduce some more notation.
Suppose $\Phi$ is a $p-$form with all its legs on $\mathcal{B},$
but with coefficients depending on $u$ (denote this by $\Phi\in\Lambda^{p}(\mathcal{B})(u)$):
\begin{equation}
\Phi=\frac{1}{p!}\Phi_{m_{1}...m_{p}}(u,x)dx^{m_{1}}\wedge...\wedge dx^{m_{p}}\ .
\end{equation}
We then introduce the restricted exterior derivative $\tilde{d}$
by 
\begin{equation}
\tilde{d}\Phi\equiv\frac{1}{\left(p+1\right)!}\left(p+1\right)\frac{\partial}{\partial x^{[n}}\Phi_{m_{1}...m_{p}]}dx^{n}\wedge dx^{m_{1}}\wedge...\wedge dx^{m_{p}}\ .
\end{equation}
We also define the operator $\mathcal{D}$ acting on such $p-$forms
as 
\begin{equation}
\mathcal{D}\Phi=\tilde{d}\Phi-\beta\wedge\dot{\Phi}\ ,
\end{equation}
where $\dot{\Phi}$ is the Lie derivative of $\Phi$ with respect
to $\partial_{u}.$ Note that
\begin{equation}
d\Phi=\mathcal{D}\Phi+He^{-}\wedge\dot{\Phi}\ .
\end{equation}
Also $\mathcal{D}$ obeys the same product rule on wedge products
as the exterior derivative and 
\begin{equation}
\mathcal{D}^{2}\Phi=-\mathcal{D}\beta\wedge\dot{\Phi}\ .
\end{equation}
Using these operators we can derive that
\begin{eqnarray}
de^{-} & = & H^{-1}\mathcal{D}\beta+e^{-}\wedge\left(H^{-1}\mathcal{D}H+\dot{\beta}\right),\nonumber \\
de^{+} & = & \mathcal{D}\omega+\frac{1}{2}\mathcal{F}\mathcal{D}\beta+He^{-}\wedge\left(\dot{\omega}+\frac{1}{2}\mathcal{F}\dot{\beta}-\frac{1}{2}\mathcal{D}\mathcal{F}\right).\label{Exterior derivative eplus and emin}
\end{eqnarray}
From these expressions it is straightforward to calculate the spin
connection components. They can be found in appendix \ref{sec:Spin-connection}.

\subsection{Supersymmetry and self-duality}

We  now derive what the conditions of section \ref{subsec:-restrictions}
become in terms of the coordinates we introduced. The three-forms
(\ref{threeforms}) become 
\begin{eqnarray}
\hat{G}^{\alpha} & = & e^{+}\wedge e^{-}\wedge\left[j^{\alpha}\left(H^{-1}\mathcal{D}H+\dot{\beta}\right)-\mathcal{D}j^{\alpha}\right]+j^{\alpha}H^{-1}e^{+}\wedge\mathcal{D}\beta\nonumber \\
 &  & +e^{-}\wedge\left[j^{\alpha}H\psi-j^{\alpha}\left(\mathcal{D}\omega\right)^{-}+H_{\mathrm{SD}}^{\alpha}\right]+\star_{4}\mathcal{D}\left(j^{\alpha}H\right)+Hj^{\alpha}\star_{4}\dot{\beta},\label{threeforms using coordinates}
\end{eqnarray}
where $\left(\mathcal{D}\omega\right)^{-}\equiv\frac{1}{2}\left(\mathcal{D}\omega-\star_{4}\mathcal{D}\omega\right),$
$\star_{4}$ is the Hodge dual on $\mathcal{B}$ and
\begin{eqnarray}
\psi & \equiv & -\frac{1}{8}H\epsilon_{abc}J_{kl}^{a}\partial_{u}\left(J^{b\ kl}\right)J^{c}.\label{psi in terms of complex structure}
\end{eqnarray}
The self-duality condition (\ref{Self-duality condition}) implies
that
\begin{eqnarray}
\mathcal{D}\beta & = & \star_{4}\mathcal{D}\beta\ ,\label{Self-duality beta}\\
H_{\mathrm{SD}}^{\alpha} & = & \star_{4}H_{\mathrm{SD}}^{\alpha}\ .\label{Self-duality two-forms}
\end{eqnarray}
Using this one can show that (\ref{+ cov derivative of two-forms})
is satisfied. The remaining constraints are the $\mu=i$ components of (\ref{gravitino condition 2})
and they constrain the covariant derivatives of $J^{a}$ on $\mathcal{B}$.
Since $d\left(e^{-}\wedge I^{a}\right)=0$ we find that
\begin{eqnarray}
\tilde{d}J^{a} & = & \mathcal{L}_{\partial_{u}}\left(\beta\wedge J^{a}\right)\ .\label{non-closure of the kahler structure}
\end{eqnarray}
Together with the fact that the $J^{a}$ satisfy the algebra (\ref{algebra kahler structure})
this implies the $\mu=i$ components of (\ref{gravitino condition 2}).
From (\ref{non-closure of the kahler structure}) we conclude that
the almost hyper-Kähler structure of $\mathcal{B}$ is not integrable
in general.

\subsection{Bianchi identity}

We can now substitute expression (\ref{threeforms using coordinates})
for the three-forms in the Bianchi identity $d\hat{G}^{\alpha}=0$.
Note that this equation is also the equation of motion of the three-forms
because of the self-duality condition. Using (\ref{Exterior derivative eplus and emin}),
the Bianchi identity reduces to

\begin{eqnarray}
\tilde{d}\left(j^{\alpha}\psi+\mathcal{G}^{+\alpha}\right) & = & \mathcal{L}_{\partial_{u}}\left[\beta\wedge\left(j^{\alpha}\psi+\mathcal{G}^{+\alpha}\right)+\star_{4}\mathcal{D}\left(j^{\alpha}H\right)+Hj^{\alpha}\star_{4}\dot{\beta}\right]\ ,\label{Bianchi identity reduced 1}
\end{eqnarray}
and
\begin{equation}
\mathcal{D}\star_{4}\left[\mathcal{D}\left(j^{\alpha}H\right)+Hj^{\alpha}\dot{\beta}\right]+\mathcal{D}\beta\wedge\mathcal{G}^{+\alpha}=0\ ,\label{Bianchi identity reduced 2}
\end{equation}
where we defined the self-dual two-forms 
\begin{equation}
\mathcal{G}^{+\alpha}=H^{-1}\left[j^{\alpha}\left(\mathcal{D}\omega\right)^{+}+\frac{1}{2}j^{\alpha}\mathcal{F}\mathcal{D}\beta+H_{\mathrm{SD}}^{\alpha}\right]\ ,
\end{equation}
with $\left(\mathcal{D}\omega\right)^{+}\equiv\frac{1}{2}\left(\mathcal{D}\omega+\star_{4}\mathcal{D}\omega\right).$

\subsection{Einstein equation}

We now consider (\ref{Einstein Equation}). In appendix \ref{sec:-component-of}
we use the spin connection components to calculate that
\begin{eqnarray}
R_{--} & = & \star_{4}\mathcal{D}\star_{4}\left(\dot{\omega}+\frac{1}{2}\mathcal{F}\dot{\beta}-\frac{1}{2}\mathcal{D}\mathcal{F}\right)-2\left(\dot{\omega}+\frac{1}{2}\mathcal{F}\dot{\beta}-\frac{1}{2}\mathcal{D}\mathcal{F}\right)^{m}\partial_{u}\left(\beta_{m}\right)\\
 &  & +\frac{1}{2}H^{-2}\left(\mathcal{D}\omega+\frac{1}{2}\mathcal{F}\mathcal{D}\beta\right)^{2}-\frac{1}{2}Hh^{mn}\partial_{u}^{2}\left(Hh_{mn}\right)-\frac{1}{4}\partial_{u}\left(Hh^{mn}\right)\partial_{u}\left(Hh_{mn}\right),\nonumber 
\end{eqnarray}
where for $\Phi_{1},\Phi_{2}\in\Lambda^{2}\left(\mathcal{B}\right)(u)$,
$\Phi_{1}\cdot\Phi_{2}=\frac{1}{2}\Phi_{1mn}\Phi_{2}^{mn}.$ From
(\ref{threeforms using coordinates}) we find that
\begin{eqnarray}
\frac{1}{4}g_{\alpha\beta}\hat{G}_{-}^{\alpha\ \rho\lambda}\hat{G}_{-\rho\lambda}^{\beta} & = & \frac{1}{2}\left[\psi-H^{-1}\left(\mathcal{D}\omega\right)^{-}\right]^{2}+\frac{1}{2}H^{-2}g_{\alpha\beta}H_{\mathrm{SD}}^{\alpha}\cdot H_{\mathrm{SD}}^{\beta}\ .
\end{eqnarray}
The scalars do not depend on $v$ so 
\begin{equation}
\partial_{-}j^{\alpha}=H\partial_{u}j^{\alpha},
\end{equation}
and the Einstein equation becomes
\begin{eqnarray}
\star_{4}\mathcal{D}\star_{4}\left(\dot{\omega}+\frac{1}{2}\mathcal{F}\dot{\beta}-\frac{1}{2}\mathcal{D}\mathcal{F}\right) & = & 2\left(\dot{\omega}+\frac{1}{2}\mathcal{F}\dot{\beta}-\frac{1}{2}\mathcal{D}\mathcal{F}\right)^{m}\partial_{u}\left(\beta_{m}\right)+\frac{1}{2}Hh^{mn}\partial_{u}^{2}\left(Hh_{mn}\right)\nonumber \\
 &  & +\frac{1}{4}\partial_{u}\left(Hh^{mn}\right)\partial_{u}\left(Hh_{mn}\right)-\frac{1}{2}H^{-2}\left(\mathcal{D}\omega+\frac{1}{2}\mathcal{F}\mathcal{D}\beta\right)^{2}\nonumber\\
 &  & +\frac{1}{2}\left[\psi-H^{-1}\left(\mathcal{D}\omega\right)^{-}\right]^{2}+\frac{1}{2}H^{-2}g_{\alpha\beta}H_{\mathrm{SD}}^{\alpha}\cdot H_{\mathrm{SD}}^{\beta}\label{Einstein equation reduced} \\
 &  & -\Omega_{\alpha\beta}H^{2}\partial_{u}j^{\alpha}\partial_{u}j^{\beta}.\nonumber 
\end{eqnarray}

\subsection{Summary}

We derived the general local form of all supersymmetric solutions
of six-dimensional $(1,0)$ supergravity with a gravity multiplet
and $n_{T}$ tensor multiplets. The metric is given by (\ref{coodinates metric})
and the three-forms by (\ref{threeforms using coordinates}). The
equations of motion can be reduced to a set of equations on the base
manifold $\mathcal{B}$. The one-form $\beta$ and two-forms $H_{\mathrm{SD}}^{\alpha}$ must obey the self-duality conditions (\ref{Self-duality beta}) and (\ref{Self-duality two-forms}). The Bianchi identity and Einstein equation reduce to (\ref{Bianchi identity reduced 1}) and
(\ref{Bianchi identity reduced 2}), and (\ref{Einstein equation reduced})
respectively. The base manifold must admit an almost hyper-Kähler
structure with almost complex structures obeying (\ref{non-closure of the kahler structure}).

\section{Classes of solutions \label{sec:Special-Cases}}

In this section we will, following \cite{Gutowski:2003rg}, consider
two cases in which the equations derived in the previous section reduce
considerably. We first focus on so-called non-twisting solutions,
which are solutions in which $\beta=0.$ After that we look at $u-$independent
solutions and consider their reduction to five dimensions. When the base-space of a $u-$independent solution is chosen to be Gibbons-Hawking and the symmetry of this metric is extended to be a symmetry of the whole solution, we show that it can be expressed in $6+2n_{T}$ harmonic functions on $\mathbb{R}^{3}$. We then briefly investigate the multi-centered Gibbons-Hawking subclass of these solutions and look at objects as the black string that have a horizon. We finish this section with some examples of pp-wave solutions.

\subsection{Non-twisting solutions\label{subsec:Non-twisting-solutions}}

Non-twisting solutions have $\beta=0$ which highly simplifies the
equations. The metric (\ref{coodinates metric}) and three-forms (\ref{threeforms using coordinates})
reduce to
\begin{equation}
ds_{6}^{2}=2H^{-1}du\left(dv+\omega+\frac{1}{2}\mathcal{F}du\right)+Hds_{4}^{2}\ ,
\end{equation}
 and
\begin{eqnarray}
\hat{G}^{\alpha} & = & -e^{+}\wedge du\wedge\tilde{d}\left(H^{-1}j^{\alpha}\right)+H^{-1}du\wedge\left[j^{\alpha}H\psi-j^{\alpha}\left(\tilde{d}\omega\right)^{-}+H_{\mathrm{SD}}^{\alpha}\right]+\star_{4}\tilde{d}\left(j^{\alpha}H\right)\ .\nonumber \\
\end{eqnarray}
The base space $\mathcal{B}$ with metric $ds_{4}^{2}$ has to be
hyper-Kähler since from (\ref{non-closure of the kahler structure})
it follows that 
\begin{equation}
\tilde{d}J^{a}=0\ .
\end{equation}
The Bianchi identity, (\ref{Bianchi identity reduced 1}) and (\ref{Bianchi identity reduced 2}),
reduce to
\begin{equation}
\tilde{d}\left(j^{\alpha}\psi+\mathcal{G}^{+\alpha}\right)=\mathcal{L}_{\partial_{u}}\star_{4}\tilde{d}\left(j^{\alpha}H\right)\ ,\label{Bianchi identity 1 non-twisting}
\end{equation}
and
\begin{equation}
\tilde{\nabla}^{2}\left(j^{\alpha}H\right)=0\ ,\label{Bianchi identity 2 non-twisting}
\end{equation}
where 
\begin{equation}
\mathcal{G}^{+\alpha}=H^{-1}\left[j^{\alpha}\left(\tilde{d}\omega\right)^{+}+H_{\mathrm{SD}}^{\alpha}\right]\ .\label{self-dual two form}
\end{equation}
Hence the functions $j^{\alpha}H$ are harmonic. The Einstein equation
(\ref{Einstein equation reduced}) reduces to
\begin{eqnarray}
\tilde{\nabla}^{i}\left(\dot{\omega}\right)_{i}-\frac{1}{2}\tilde{\nabla}^{2}\mathcal{F} & = & \frac{1}{2}Hh^{mn}\partial_{u}^{2}\left(Hh_{mn}\right)+\frac{1}{4}\partial_{u}\left(Hh^{mn}\right)\partial_{u}\left(Hh_{mn}\right)-\frac{1}{2}H^{-2}\left(\tilde{d}\omega\right)^{2}\label{Einstein equation non-twisting}\\
 &  & +\frac{1}{2}\left[\psi-H^{-1}\left(\tilde{d}\omega\right)^{-}\right]^{2}+\frac{1}{2}H^{-2}g_{\alpha\beta}H_{\mathrm{SD}}^{\alpha}\cdot H_{\mathrm{SD}}^{\beta}-\Omega_{\alpha\beta}H^{2}\partial_{u}j^{\alpha}\partial_{u}j^{\beta}.\nonumber 
\end{eqnarray}
In principle one should be able to solve these equations successively.
First pick a hyper-Kähler base space $\mathcal{B}$ and pick harmonic
functions on this space for $j^{\alpha}H$. The function $H$ then
follows by using the identity $j_{\alpha}j^{\alpha}=1.$ The two-form
$\mathcal{G}^{+\alpha}$ can be determined by using its self-duality
and (\ref{Bianchi identity 1 non-twisting}). Then $\omega$ can be
determined by contracting (\ref{self-dual two form}) with $j_{\alpha}$
which then also gives an expression for $H_{\mathrm{SD}}^{\alpha}.$
Lastly, $\mathcal{F}$ can be determined from (\ref{Einstein equation non-twisting}).

Solutions that are dependent on $u$ (but not necessarily non-twisting)
have been studied in the case of minimal supergravity or in the case
with an extra tensor multiplet \cite{Ford:2006yb,Bobev:2012af,Lunin:2012gp,Giusto:2013rxa,Shigemori:2013lta}.
To show that one can still do this with an arbitrary number of tensor
multiplets, we construct an explicit example of a $u-$dependent solution.

\paragraph*{Flat base space.}

We will extend the non-twisting solution of \cite{Gutowski:2003rg}
with flat base space to a solution with tensor multiplets. As metric
on the base space we take
\begin{equation}
ds_{4}^{2}=dr^{2}+\frac{1}{4}r^{2}\left[\left(\sigma^{1}\right)^{2}+\left(\sigma^{2}\right)^{2}+\left(\sigma^{3}\right)^{2}\right],
\end{equation}
where $\sigma^{a}, a=1,2,3,$ are either the left-invariant $\sigma_{R}^{a}$
or the right-invariant $\sigma_{L}^{a}$ one-forms on the three-sphere:
$d\sigma^{a}=\frac{1}{2}\eta\epsilon_{\ \,bc}^{a}\sigma^{b}\wedge\sigma^{c}$
with $\eta=1$ if $\sigma=\sigma_{R}$ and $\eta=-1$ if $\sigma=\sigma_{L}.$
We can then take the vierbein
\begin{eqnarray}
\tilde{e}^{4} & = & dr,\nonumber \\
\tilde{e}^{a} & = & \frac{r}{2}\sigma^{a}.
\end{eqnarray}
If we take the hyper-Kähler structure (in Cartesian coordinates) given by
\begin{eqnarray}
J^{1} & = & -\left(dx^{1} \wedge dx^{3}+dx^{2} \wedge dx^{4}\right),\nonumber \\
J^{2} & = & -\left(dx^{1} \wedge dx^{2}-dx^{3} \wedge dx^{4}\right),\label{hyper structure flat space}\\
J^{3} & = & -\left(dx^{1} \wedge dx^{4}-dx^{2} \wedge dx^{3}\right),\nonumber 
\end{eqnarray}
we have that $\psi=0.$ See \cite{Gauntlett:2002nw} for the coordinate transformation to express these forms in terms of $\sigma^{a}$. Requiring the two-forms (\ref{hyper structure flat space}) to be anti-self-dual imposes the orientation $\tilde{e}^{4}\wedge\tilde{e}^{1}\wedge\tilde{e}^{2}\wedge\tilde{e}^{3}$. For the
simple case that $j^{\alpha}$ and $H$ only depend on $u$ and $r,$
we can write the harmonic functions 
\begin{equation}
j^{\alpha}H=P^{\alpha}(u)+\frac{Q^{\alpha}(u)}{r^{2}},\label{scalars non-twisting flat base-space}
\end{equation}
where $P^{\alpha},$ $Q^{\alpha}$ are arbitrary functions of $u$
that will be fixed by the other equations of motion. From $j_{\alpha}j^{\alpha}=1$
we find
\begin{equation}
H=\sqrt{\Omega_{\alpha\beta}\left(P^{\alpha}+\frac{Q^{\alpha}}{r^{2}}\right)\left(P^{\beta}+\frac{Q^{\beta}}{r^{2}}\right)}\ .\label{function H non-twisting flat base-space}
\end{equation}
Notice that (\ref{Bianchi identity 1 non-twisting}) reduces to
\begin{equation}
\tilde{d}\mathcal{G}^{+\alpha}=\mathcal{L}_{\partial_{u}}\star_{4}\tilde{d}\left(j^{\alpha}H\right).\label{bianchi identity part 1 non-twisting flat base space}
\end{equation}
Using the self-duality of $\mathcal{G}^{+\alpha}$ we can write 
\begin{equation}
\mathcal{G}^{+\alpha}=C_{b}^{\alpha}\tilde{e}^{4}\wedge\tilde{e}^{b}+\frac{1}{2}C_{b}^{\alpha}\epsilon_{\ cd}^{b}\tilde{e}^{c}\wedge\tilde{e}^{d},
\end{equation}
where we assume that $C_{b}^{\alpha}$ only depend on $u$ and $r$
(to stay in line with \cite{Gutowski:2003rg}). We can calculate 
\begin{eqnarray}
\tilde{d}\mathcal{G}^{+\alpha} & = & \frac{1}{r^{2}}\left[\left(1-\eta\right)rC_{b}^{\alpha}+\frac{1}{2}r^{2}\partial_{r}\left(C_{b}^{\alpha}\right)\right]\epsilon_{\ cd}^{b}\tilde{e}^{4}\wedge\tilde{e}^{c}\wedge\tilde{e}^{d}\ .
\end{eqnarray}
Substitution in (\ref{bianchi identity part 1 non-twisting flat base space})
yields 
\begin{eqnarray}
\frac{1}{r^{2}}\left[\left(1-\eta\right)rC_{b}^{\alpha}+\frac{1}{2}r^{2}\partial_{r}\left(C_{b}^{\alpha}\right)\right]\epsilon_{\ cd}^{b}\tilde{e}^{4}\wedge\tilde{e}^{c}\wedge\tilde{e}^{d} & = & \mathcal{L}_{\partial_{u}}\star_{4}\tilde{d}\left(j^{\alpha}H\right)\nonumber \\
 & = & -2\frac{\partial_{u}Q^{\alpha}}{r^{3}}\tilde{e}^{1}\wedge\tilde{e}^{2}\wedge\tilde{e}^{3},\label{Right hand side bianchi identity}
\end{eqnarray}
from which it follows that
\begin{eqnarray}
\partial_{u}Q^{\alpha} & = & 0\ ,\nonumber \\
\partial_{r}\left(C_{b}^{\alpha}\right) & = & 2(\eta-1)\frac{1}{r}C_{b}^{\alpha}\ .
\end{eqnarray}
The second equation is solved by
\begin{equation}
C_{b}^{\alpha}=C_{b}^{\alpha}(u)r^{2(\eta-1)}\ ,
\end{equation}
for functions $C_{b}^{\alpha}(u).$ Hence 
\begin{equation}
\mathcal{G}^{+\alpha}=C_{b}^{\alpha}(u)r^{2(\eta-1)}\tilde{e}^{4}\wedge\tilde{e}^{b}+\frac{1}{2}C_{b}^{\alpha}(u)r^{2(\eta-1)}\epsilon_{\ cd}^{b}\tilde{e}^{c}\wedge\tilde{e}^{d}\ .\label{self-dual G flat base-space nontwisting}
\end{equation}
Assuming as in \cite{Gutowski:2003rg} that 
\begin{equation}
\omega=W(u,r)\sigma^{3},\label{omega non-twisting flat base-space}
\end{equation}
we can calculate that 
\begin{equation}
\left(\tilde{d}\omega\right)^{+}=\left(\frac{2}{r^{2}}\eta W+\frac{1}{r}\partial_{r}W\right)\left(\tilde{e}^{1}\wedge\tilde{e}^{2}-\tilde{e}^{3}\wedge\tilde{e}^{4}\right).\label{omega plus non-twisting flat base-space}
\end{equation}
We then find from substitution of (\ref{scalars non-twisting flat base-space}),
(\ref{self-dual G flat base-space nontwisting}) and (\ref{omega plus non-twisting flat base-space})
in 
\begin{equation}
\left(\tilde{d}\omega\right)^{+}=H j_{\alpha}\mathcal{G}^{+\alpha} \label{differential omega}\ ,
\end{equation}
that
\begin{eqnarray}
C_{1}^{\alpha}(u) & = & C_{2}^{\alpha}(u)=0\ ,\nonumber \\
W & = & W_{1}(u)r^{-2\eta}+\frac{1}{2}\Omega_{\alpha\beta}C_{3}^{\alpha}(u)\left(\frac{P^{\beta}}{2\eta}r^{2\eta}+\frac{Q^{\beta}}{2\eta-1}r^{2\eta-2}\right),\label{solving for W non-twisting flat base-space}
\end{eqnarray}
where $W_{1}$ is yet another arbitrary function of $u$. We then
find from (\ref{scalars non-twisting flat base-space}), (\ref{self-dual G flat base-space nontwisting}) and (\ref{differential omega}) that 
\begin{eqnarray}
H_{\mathrm{SD}}^{\alpha} & = & H\mathcal{G}^{+\alpha}-j^{\alpha}\left(\tilde{d}\omega\right)^{+}\nonumber \\
 & = & \left[HC_{3}^{\alpha}(u)r^{2(\eta-1)}-j^{\alpha}C_{3}^{\beta}(u)r^{2(\eta-1)}\Omega_{\beta\gamma}\left(P^{\gamma}(u)+\frac{Q^{\gamma}(u)}{r^{2}}\right)\right]\left(\tilde{e}^{1}\wedge\tilde{e}^{2}-\tilde{e}^{3}\wedge\tilde{e}^{4}\right).\nonumber \\
\label{self-dual tensors non-twisting flat base-space}
\end{eqnarray}
Lastly, the Einstein equation (\ref{Einstein equation non-twisting})
reduces to 
\begin{eqnarray}
\tilde{\nabla}^{i}\left(\dot{\omega}\right)_{i}-\frac{1}{2}\tilde{\nabla}^{2}\mathcal{F} & = & 2H\partial_{u}^{2}\left(H\right)+\partial_{u}\left(H\right)\partial_{u}\left(H\right)-\frac{1}{2}H^{-2}\left(\left(\tilde{d}\omega\right)^{+}\right)^{2}\nonumber \\
 &  & -\frac{1}{2}H^{-2}\Omega_{\alpha\beta}H_{\mathrm{SD}}^{\alpha}\cdot H_{\mathrm{SD}}^{\beta}-\Omega_{\alpha\beta}H^{2}\partial_{u}j^{\alpha}\partial_{u}j^{\beta}.
\end{eqnarray}
Using (\ref{scalars non-twisting flat base-space}), (\ref{function H non-twisting flat base-space}),
(\ref{omega non-twisting flat base-space}), (\ref{solving for W non-twisting flat base-space})
and (\ref{self-dual tensors non-twisting flat base-space}), and assuming
that $\mathcal{F}=\mathcal{F}(u,r)$ we derive that
\begin{equation}
\partial_{r}\left(r^{3}\partial_{r}\mathcal{F}\right)=-2\Omega_{\alpha\beta}\dot{P}^{\alpha}\dot{P}^{\beta}r^{3}-4\Omega_{\alpha\beta}\left(P^{\alpha}r^{3}+Q^{\alpha}r\right)\partial_{u}^{2}P^{\beta}+2\Omega_{\alpha\beta}C_{3}^{\alpha}(u)C_{3}^{\beta}(u)r^{4\eta-1}.
\end{equation}
Integration of this equation yields
\begin{eqnarray}
\mathcal{F} & = & C_{5}(u)-\frac{1}{2}C_{4}(u)\frac{1}{r^{2}}-\frac{1}{2}\Omega_{\alpha\beta}\left(P^{\alpha}\partial_{u}^{2}P^{\beta}+\frac{1}{2}\dot{P}^{\alpha}\dot{P}^{\beta}\right)r^{2}\nonumber \\
 &  & +\frac{1}{4\eta\left(2\eta-1\right)}\Omega_{\alpha\beta}C_{3}^{\alpha}(u)C_{3}^{\beta}(u)r^{4\eta-2}-2\Omega_{\alpha\beta}Q^{\alpha}\partial_{u}^{2}P^{\beta}\log(r).
\end{eqnarray}
for arbitrary functions $C_{4}$ and $C_{5}$. This construction can
easily be extended to other hyper-Kähler base spaces. 

\subsection{$u-$independent solutions}

A second class of solutions in which the general equations simplify considerably is the class that does not depend on $u.$ We can see this as introducing an extra
symmetry of the solution. In particular we get an extra Killing vector
$\partial_{u},$ which is spacelike when $\mathcal{F}>0$ and timelike
when $\mathcal{F}<0.$ The three-forms reduce to 
\begin{eqnarray}
\hat{G}^{\alpha} & = & -He^{+}\wedge e^{-}\wedge\tilde{d}\left(H^{-1}j^{\alpha}\right)+j^{\alpha}H^{-1}e^{+}\wedge\tilde{d}\beta+e^{-}\wedge\left[-j^{\alpha}\left(\tilde{d}\omega\right)^{-}+H_{\mathrm{SD}}^{\alpha}\right]+\star_{4}\tilde{d}\left(j^{\alpha}H\right).\nonumber \\ \label{threeforms u-independent solutions}
\end{eqnarray}
The base space has to be hyper-Kähler since from (\ref{non-closure of the kahler structure})
\begin{equation}
\tilde{d}J^{a}=0\ ,\label{u-indep integrability hyper-kahler}
\end{equation}
and $\beta$ has self-dual curvature:
\begin{equation}
\tilde{d}\beta=\star_{4}\tilde{d}\beta\ .\label{u-indep self-duality}
\end{equation}
The Bianchi identity, (\ref{Bianchi identity reduced 1}) and (\ref{Bianchi identity reduced 2}),
and Einstein equation (\ref{Einstein equation reduced}) reduce to
respectively
\begin{eqnarray}
\tilde{d}\mathcal{G}^{+\alpha} & = & 0\ ,\label{u-indep Bianchi identity 1}\\
\tilde{d}\star_{4}\tilde{d}\left(j^{\alpha}H\right) & = & -\tilde{d}\beta\wedge\mathcal{G}^{+\alpha}\ ,\label{u-indep Bianchi identity 2}\\
\tilde{\nabla}^{2}\mathcal{F} & = & \Omega_{\alpha\beta}\mathcal{G}^{+\alpha}\cdot\mathcal{G}^{+\beta}\ ,\label{u-indep Einstein Eq.}
\end{eqnarray}
where
\begin{equation}
\mathcal{G}^{+\alpha}=H^{-1}\left[j^{\alpha}\left(\tilde{d}\omega\right)^{+}+\frac{1}{2}j^{\alpha}\mathcal{F}\tilde{d}\beta+H_{\mathrm{SD}}^{\alpha}\right]\ ,
\end{equation}
and we have used that 
\begin{eqnarray}
\Omega_{\alpha\beta}\mathcal{G}^{+\alpha}\cdot\mathcal{G}^{+\beta} & = & H^{-2}\left[\left(\tilde{d}\omega\right)^{+}+\frac{1}{2}\mathcal{F}\tilde{d}\beta\right]\cdot\left[\left(\tilde{d}\omega\right)^{+}+\frac{1}{2}\mathcal{F}\tilde{d}\beta\right]-H^{-2}g_{\alpha\beta}H_{\mathrm{SD}}^{\alpha}\cdot H_{\mathrm{SD}}^{\beta}\ .\nonumber \\
\end{eqnarray}
When the Killing vector $\partial_{u}$ is spacelike, the $u-$direction
can be compactified on a circle and we can reduce the solution to
five dimensions. This will be done in the next section. In section \ref{subsec:Gibbons-Hawking-base-space}
we then take the base space $\mathcal{B}$ to be Gibbons-Hawking \cite{Gibbons:1979zt}
which has yet another Killing vector. Assuming that this symmetry
extends to the whole solution, we solve the equations of motion completely. 

\subsection{Reduction to five dimensions \label{subsec:Reduction-to-five-dimensions}}

When one considers a $u-$independent solution with $\mathcal{F}$
positive such that $\partial_{u}$ is a spacelike Killing vector,
one can compactify this direction on a circle and do a Kaluza-Klein
reduction to obtain a five-dimensional solution. The six-dimensional
metric reduces to the five-dimensional metric $ds_{5}^{2}$, a Kaluza-Klein
vector $A^{0}$ and a scalar $X^{0}.$ The three-forms $\hat{G}^{\alpha}$
reduce to three-forms $G^{\alpha}$ and two-forms $F^{\alpha}$ that
are related to each other since the $\hat{G}^{\alpha}$ obey a self-duality
condition. The scalars $j^{\alpha}$ reduce to scalars $X^{\alpha}.$
Reducing the six-dimensional theory to five-dimensions thus results
in five-dimensional  supergravity coupled to $n_{T}+1$ vector multiplets.
We can express the six-dimensional data in terms of five-dimensional
data by \cite{Bonetti:2011mw} 
\begin{eqnarray}
ds_{6}^{2} & = & r^{2}\left(du+A^{0}\right)^{2}+r^{-2/3}ds_{5}^{2}\ ,\nonumber \\
\hat{G}^{\alpha} & = & G^{\alpha}-F^{\alpha}\wedge\left(du+A^{0}\right)\ ,\label{ansatz three-forms}\\
j^{\alpha} & = & r^{-2/3}X^{\alpha}\ ,\label{ansatz scalars}\\
r^{-4/3} & = & X^{0}\ ,\nonumber 
\end{eqnarray}
where 
\begin{eqnarray}
ds_{5}^{2} & = & -f^{2}\left(dv+\omega\right)^{2}+f^{-1}ds_{4}^{2}\ ,\nonumber \\
G^{\alpha} & = & dB^{\alpha}+A^{\alpha}\wedge F^{0}\ ,
\end{eqnarray}
for a function $f$ and two-forms $B^{\alpha}$. The scalars $X^{I},$ $I\in\{0,1,...,n_{T}+1\},$
are the so-called very special coordinates. These are are $n_{T}+2$
real coordinates that describe an $n_{T}+2-$dimensional manifold
in which the scalar manifold is given by the hypersurface \cite{Gunaydin:1983bi}
\begin{equation}
\mathcal{N}\equiv\frac{1}{3!}C_{IJK}X^{I}X^{J}X^{K}=1\ ,\label{Cubic potential}
\end{equation}
where $C_{IJK}$ is a constant symmetric tensor and $\mathcal{N}$
is the so-called cubic potential. The potential $\mathcal{N}$ in
terms of six-dimensional data is given by \cite{Bonetti:2011mw}
\begin{equation}
\mathcal{N}=\Omega_{\alpha\beta}X^{0}X^{\alpha}X^{\beta}\ .\label{cubic potential in 6d data}
\end{equation}
It is straightforward to derive that 
\begin{eqnarray}
ds_{5}^{2} & = & -\left(\mathcal{F}H^{2}\right)^{-2/3}\left(dv+\omega\right)^{2}+\left(\mathcal{F}H^{2}\right)^{1/3}ds_{4}^{2}\ ,\nonumber \\
X^{0} & = & \left(\mathcal{F}H^{-1}\right)^{-2/3}\ ,\nonumber \\
X^{\alpha} & = & \left(\mathcal{F}H^{-1}\right)^{1/3}j^{\alpha}\ ,\\
F^{0} & = & \tilde{d}\left[\beta+\mathcal{F}^{-1}\left(dv+\omega\right)\right]\ ,\nonumber \\
F^{\alpha} & = & \tilde{d}\left[H^{-1}j^{\alpha}\left(dv+\omega\right)\right]-H^{-1}\left[j^{\alpha}\left(\tilde{d}\omega\right)^{+}+\frac{1}{2}j^{\alpha}\mathcal{F}\tilde{d}\beta+H_{\mathrm{SD}}^{\alpha}\right]\ .\nonumber 
\end{eqnarray}
Note that the field strengths can be written as 
\begin{eqnarray}
F^{I} & = & \tilde{d}\left[X^{I}f\left(dv+\omega\right)\right]+\Theta^{I}\ ,
\end{eqnarray}
where 
\begin{equation}
\Theta^{0}  =  \tilde{d}\beta\ ,\qquad \Theta^{\alpha}  =  -\mathcal{G}^{\alpha}\ ,\label{Self-dual tensors 4d}
\end{equation}
are self-dual tensors. Also (\ref{u-indep Bianchi identity 1}) implies
that the two-forms $\Theta^{I}$ are closed. Using 
\begin{equation}
X_{I}\equiv\frac{1}{6}C_{IJK}X^{J}X^{K}\ ,
\end{equation}
we find that
\begin{eqnarray}
X_{I}\Theta^{I} & = & -\frac{2}{3}f\left(\tilde{d}\omega\right)^{+}.
\end{eqnarray}
Furthermore, we find that (\ref{u-indep Bianchi identity 2}) reduces
to 
\begin{equation}
\tilde{\nabla}^{2}\left(f^{-1}X_{\alpha}\right)=\frac{1}{6}C_{\alpha JK}\Theta^{J}\cdot\Theta^{K}\ ,
\end{equation}
and that the Einstein equation (\ref{u-indep Einstein Eq.}) reduces
to 
\begin{equation}
\tilde{\nabla}^{2}\left(f^{-1}X_{0}\right)=\frac{1}{6}C_{0JK}\Theta^{J}\cdot\Theta^{K}\ .\label{eq: reduced Einstein equation}
\end{equation}
This implies that we find ourselves exactly in the timelike class
of five-dimensional solutions of \cite{Gutowski:2004yv,Gauntlett:2004qy}. Thus every
solution of  five-dimensional supergravity coupled to an arbitrary
number of vector multiplets in the timelike class that has a cubic
potential of the form
\begin{equation}
\mathcal{N}=\frac{1}{3!}C_{IJK}X^{I}X^{J}X^{K}=\Omega_{\alpha\beta}X^{0}X^{\alpha}X^{\beta}\ ,
\end{equation}
can be uplifted to six dimensions.\textbf{ }An interesting remark
is that classical $M-$theory solutions do not have cubic potentials
of this form. To lift them up, one also has to take into account the
one-loop contributions coming from the reduction on the circle \cite{Bonetti:2011mw,Bonetti:2013ela}.

\paragraph{Minimal five-dimensional supergravity.}

After the reduction we always end up with at least one vector multiplet
in five dimensions. However, one can truncate the reduction of minimal
supergravity in six dimensions to minimal five-dimensional supergravity
\cite{Gutowski:2003rg}. We get minimal supergravity when we set the
three-forms of the tensor multiplets $H^{M}=0$ for $M=1,...,n_{T}$
and furthermore set $j^{\alpha}=0$ for $\alpha=2,...,n_{T}+1$ and
$j^{1}=1.$ The only three-form that is non-zero is $\hat{G}^{1}.$
We can truncate the reduced theory to minimal supergravity by getting
rid of the scalars, which can be done by setting $\mathcal{F}=H$
such that $X^{0}=X^{1}=1.$ Consistency of (\ref{u-indep Bianchi identity 2})
and (\ref{u-indep Einstein Eq.}) then enforces $\tilde{d}\beta=-\mathcal{G}^{+1}$
or
\begin{equation}
\tilde{d}\beta=-\frac{2}{3}H^{-1}\left(\tilde{d}\omega\right)^{+}.
\end{equation}
This implies that 
\begin{eqnarray}
F\equiv F^{0}=F^{1} & = & \tilde{d}\left[\beta+\mathcal{F}^{-1}\left(dv+\omega\right)\right].
\end{eqnarray}
Introducing $G^{+}\equiv f\left(\tilde{d}\omega\right)^{+},$ we find
that 
\begin{equation}
\tilde{d}G^{+}=0
\end{equation}
and that (\ref{eq: reduced Einstein equation}) reduces to
\begin{equation}
\tilde{\nabla}^{2}\left(f^{-1}\right)=\frac{4}{9}G^{+}\cdot G^{+}.
\end{equation}
This means that we find ourselves in the timelike class of minimal
five-dimensional supergravity \cite{Gauntlett:2002nw}\footnote{They use a different normalization of the field strength: $F_{\mathrm{here}}=\frac{2}{\sqrt{3}}F_{\mathrm{there}}.$}.

The null class of minimal five-dimensional supergravity arises from
reducing non-twisting solutions of minimal six-dimensional supergravity
that have a Gibbons-Hawking base space \cite{Gutowski:2003rg}. 

\subsection{Gibbons-Hawking base space\label{subsec:Gibbons-Hawking-base-space}}

We now consider $u-$independent solutions with a Gibbons-Hawking
base space \cite{Gibbons:1979zt}. This is the most general hyper-Kähler
four-manifold admitting a Killing vector field $\partial_{\psi}$\footnote{Notice that this $\psi$ is not related to the $\psi$ in terms of
the complex structures. We can safely do this since from (\ref{psi in terms of complex structure})
we see that $\psi=0$ for this class of solutions.} preserving the three complex structures \cite{Gibbons:1987sp}. It
has the metric

\begin{equation}
ds_{4}^{2}=H_{2}^{-1}\left(d\psi+\chi_{a}dx^{a}\right)^{2}+H_{2}\delta_{ab}dx^{a}dx^{b},
\end{equation}
where $a=1,2,3,$ $\chi_{a}$ and $H_{2}$ are independent of $\psi$
and
\begin{eqnarray}
\nabla^{2}H_{2} & = & 0,\nonumber \\
\vec{\nabla}\times\vec{\chi} & = & \vec{\nabla}H_{2}.
\end{eqnarray}
We take $\nabla$ with respect to three-dimensional flat space. 

We now obtain all solutions in the case the symmetry $\partial_{\psi}$
of the base space is extended to a symmetry of the full solution.
This was done in \cite{Gauntlett:2002nw} for minimal five-dimensional
supergravity, in \cite{Gutowski:2003rg} for minimal six-dimensional
supergravity and in \cite{Gauntlett:2004qy} for  five-dimensional
supergravity coupled to an arbitrary number of vector multiplets,
so we will be quite brief here.

We can choose the vierbein
\begin{eqnarray}
\tilde{e}^{4} & = & H_{2}^{-1/2}\left(d\psi+\chi_{a}dx^{a}\right),\nonumber \\
\tilde{e}^{a} & = & H_{2}^{1/2}dx^{a}.
\end{eqnarray}
Anti-self-duality of the complex structure forms implies that the
volume form is given by $\tilde{e}^{4}\wedge\tilde{e}^{1}\wedge\tilde{e}^{2}\wedge\tilde{e}^{3}.$
We can decompose the one-forms
\begin{eqnarray}
\beta & = & \beta_{0}\left(d\psi+\chi_{a}dx^{a}\right)+\beta_{a}dx^{a},\nonumber \\
\omega & = & \omega_{0}\left(d\psi+\chi_{a}dx^{a}\right)+\omega_{a}dx^{a},
\end{eqnarray}
where $\beta_{0},$ $\beta_{a},$ $\omega_{0}$ and $\omega_{a}$
are functions on $\mathbb{R}^{3}.$ Solving (\ref{u-indep self-duality})
results in

\begin{equation}
\beta_{0}  =  H_{2}^{-1}H_{3}\ ,\qquad
\vec{\nabla}\times\vec{\beta}  =  -\vec{\nabla}H_{3}\ ,
\end{equation}
with $H_{3}$ an arbitrary harmonic function on $\mathbb{R}^{3}$.
The two-form $\mathcal{G}^{+\alpha}$ is self-dual so it has to be
of the form
\begin{equation}
\mathcal{G}^{+\alpha}=-\frac{1}{2}C_{b}^{\alpha}\tilde{e}^{4}\wedge\tilde{e}^{b}-\frac{1}{4}C_{b}^{\alpha}\epsilon_{\ cd}^{b}\tilde{e}^{c}\wedge\tilde{e}^{d},
\end{equation}
and solving (\ref{u-indep Bianchi identity 1}) results in
\begin{eqnarray}
\vec{C}^{\alpha} & = & 2\vec{\nabla}\left(H_{2}^{-1}H_{4}^{\alpha}\right)\ ,
\end{eqnarray}
with $H_{4}^{\alpha}$ arbitrary harmonic functions on $\mathbb{R}^{3}.$
Using this result one can solve (\ref{u-indep Bianchi identity 2}),
which results in 
\begin{equation}
j^{\alpha}H=H_{1}^{\alpha}-H_{2}^{-1}H_{3}H_{4}^{\alpha}\ ,\label{result scalars}
\end{equation}
where $H_{1}^{\alpha}$ are arbitrary harmonic functions on $\mathbb{R}^{3}$.
Using $j_{\alpha}j^{\alpha}=1$ we find
\begin{equation}
H=\sqrt{\Omega_{\alpha\beta}\left(H_{1}^{\alpha}-H_{2}^{-1}H_{3}H_{4}^{\alpha}\right)\left(H_{1}^{\beta}-H_{2}^{-1}H_{3}H_{4}^{\beta}\right)}\ .\label{Result H}
\end{equation}
With the solution of $\mathcal{G}^{+\alpha},$ (\ref{u-indep Einstein Eq.})
can be solved and yields
\begin{equation}
\mathcal{F}=-H_{5}+H_{2}^{-1}\Omega_{\alpha\beta}H_{4}^{\alpha}H_{4}^{\beta}\ ,\label{Result F}
\end{equation}
with $H_{5}$ an arbitrary harmonic function on $\mathbb{R}^{3}.$
Now, using that 
\begin{eqnarray}
j_{\alpha}\mathcal{G}^{+\alpha} & = & -j_{\alpha}\nabla_{b}\left(H_{2}^{-1}H_{4}^{\alpha}\right)\tilde{e}^{4}\wedge\tilde{e}^{b}-\frac{1}{2}j_{\alpha}\nabla_{b}\left(H_{2}^{-1}H_{4}^{\alpha}\right)\epsilon_{\ cd}^{b}\tilde{e}^{c}\wedge\tilde{e}^{d}\ ,\nonumber \\
 & = & H^{-1}\left(\tilde{d}\omega\right)^{+}+\frac{1}{2}\mathcal{F}H^{-1}\tilde{d}\beta\ ,\label{contraction self-dual G}
\end{eqnarray}
we get an equation for $\omega:$ 
\begin{eqnarray}
H_{2}\vec{\nabla}\omega_{0}-\omega_{0}\vec{\nabla}H_{2}-\vec{\nabla}\times\vec{\omega} & = & 2\Omega_{\alpha\beta}\left(H_{1}^{\alpha}H_{2}-H_{3}H_{4}^{\alpha}\right)\vec{\nabla}\left(H_{2}^{-1}H_{4}^{\beta}\right)\nonumber \\
 &  & +\left(H_{5}H_{2}-\Omega_{\alpha\beta}H_{4}^{\alpha}H_{4}^{\beta}\right)\vec{\nabla}\left(H_{2}^{-1}H_{3}\right).\label{eq:equation for omega}
\end{eqnarray}
Taking the divergence of this equation yields an integrability equation
that can be solved for $\omega_{0}:$ 
\begin{equation}
\omega_{0}=H_{6}-H_{2}^{-2}H_{3}\Omega_{\alpha\beta}H_{4}^{\alpha}H_{4}^{\beta}+H_{2}^{-1}\Omega_{\alpha\beta}H_{1}^{\alpha}H_{4}^{\beta}+\frac{1}{2}H_{2}^{-1}H_{3}H_{5},
\end{equation}
with $H_{6}$ an arbitrary harmonic function on $\mathbb{R}^{3}.$
Substitution of this in (\ref{eq:equation for omega}) gives an equation
that determines $\vec{\omega}$ up to a gradient (and this can be
eliminated by shifting $v$):
\begin{eqnarray}
\vec{\nabla}\times\vec{\omega} & = & \Omega_{\alpha\beta}\left(H_{4}^{\alpha}\vec{\nabla}H_{1}^{\beta}-H_{1}^{\alpha}\vec{\nabla}H_{4}^{\beta}\right)+H_{2}\vec{\nabla}H_{6}-H_{6}\vec{\nabla}H_{2}+\frac{1}{2}H_{3}\vec{\nabla}H_{5}-\frac{1}{2}H_{5}\vec{\nabla}H_{3}.\nonumber \\ \label{curl omega 1}
\end{eqnarray}
From the definition of $\mathcal{G}^{+\alpha}$ we then find that
\begin{equation}
H_{\mathrm{SD}}^{\alpha}=-D_{b}^{\alpha}\tilde{e}^{4}\wedge\tilde{e}^{b}-\frac{1}{2}D_{b}^{\alpha}\epsilon_{\ cd}^{b}\tilde{e}^{c}\wedge\tilde{e}^{d},
\end{equation}
where
\begin{equation}
\vec{D}^{\alpha}\equiv H\vec{\nabla}\left(H_{2}^{-1}H_{4}^{\alpha}\right)-j^{\alpha}Hj_{\beta}\vec{\nabla}\left(H_{2}^{-1}H_{4}^{\beta}\right).
\end{equation}

We now consider the so-called multi-centered Gibbons-Hawking subclass of these solutions. We introduce the notation
\begin{equation}
\mathbb{H}\equiv(H_{1}^{\alpha},H_{2},H_{3},H_{4}^{\alpha},H_{5},H_{6}),\ \Gamma_{A}\equiv(\mu_{A}^{\alpha},m_{A},q_{A},p_{A}^{\alpha},n_{A},j_{A}),\ \Gamma_{\infty}\equiv(\mu_{\infty}^{\alpha},m_{\infty},q_{\infty},p_{\infty}^{\alpha},n_{\infty},j_{\infty}),
\end{equation}
where $A=1,...,N$ and all the components
of the vectors $\Gamma_{A}$ and $\Gamma_{\infty}$ are constants.
We then take the harmonic functions of the form
\begin{eqnarray}
\mathbb{H} & = & \Gamma_{\infty}+\sum_{A}\frac{\Gamma_{A}}{|\vec{x}-\vec{x}_{A}|}.\label{form harmonic functions}
\end{eqnarray}
Although every set of centers $\vec{x}_{A}$ describes a solution,
there will typically be Dirac string-like singularities. Imposing
the absence of these singularities gives a constraint on the relative
positions, see \cite{Bena:2005va,Bena:2008wt,Denef:2000nb}. This
arises from requiring $\vec{\omega}$ to be globally defined, which
implies that $d^{2}\vec{\omega}=0$. If we define the symplectic product
$\langle \rangle$ working on vectors of the form $v=(v_{1}^{\alpha},v_{2},v_{3},v_{4}^{\alpha},v_{5},v_{6})$
via
\begin{equation}
\langle v,w \rangle=\Omega_{\alpha\beta}v_{4}^{\alpha}w_{1}^{\beta}-\Omega_{\alpha\beta}v_{1}^{\alpha}w_{4}^{\beta}+v_{2}w_{6}-v_{6}w_{2}+\frac{1}{2}\left(v_{3}w_{5}-v_{5}w_{3}\right),
\end{equation}
we can rewrite (\ref{curl omega 1}) as
\begin{equation}
\star_{3}d\vec{\omega}=\langle \mathbb{H},d\mathbb{H} \rangle.
\end{equation}
Taking $d\star_{3}$ on both sides leads to 
\begin{equation}
\sum_{B\neq A}\frac{\langle \Gamma_{A},\Gamma_{B} \rangle}{|\vec{x}_{A}-\vec{x}_{B}|}=\langle \Gamma_{\infty},\Gamma_{A}\rangle,\ \ \ A=1,...,N.\label{Bubble equations}
\end{equation}
These are usually referred to as ``Bubble equations'' since they
control the size of the two-cycles or bubbles in the Gibbons-Hawking
base space \cite{Bena:2007kg}.

Let $Sp\left(6+2n_{T},\mathbb{R}\right)$ denote the symplectic group
that preserves the symplectic product $\langle \rangle.$ A linear combination
of harmonic functions is still harmonic, hence sending $\mathbb{H}\rightarrow g\mathbb{H}$
with $g\in Sp\left(6+2n_{T},\mathbb{R}\right)$ sends a solution to
a solution, preserving regularity. This symplectic group was earlier noticed
for minimal supergravity in \cite{Crichigno:2016lac}.

\paragraph*{Summary.}

The most general $u-$independent solution with a Gibbons-Hawking
base space whose Killing vector field extends to a symmetry of the
full solution is determined by $6+2n_{T}$ harmonic functions $H_{1}^{\alpha},$
$H_{2},$ $H_{3},$ $H_{4}^{\alpha},$ $H_{5}$ and $H_{6}$ on $\mathbb{R}^{3}.$
Its metric is given by
\begin{eqnarray}
ds_{6}^{2} & = & 2H^{-1}\left(du+\beta\right)\left(dv+\omega+\frac{1}{2}\mathcal{F}\left(du+\beta\right)\right)+Hds_{4}^{2},\nonumber \\
ds_{4}^{2} & = & H_{2}^{-1}\left(d\psi+\chi_{a}dx^{a}\right)^{2}+H_{2}\delta_{ab}dx^{a}dx^{b},
\end{eqnarray}
where
\begin{eqnarray}
\vec{\nabla}\times\vec{\chi} & = & \vec{\nabla}H_{2},\nonumber \\
H & = & \sqrt{\Omega_{\alpha\beta}\left(H_{1}^{\alpha}-H_{2}^{-1}H_{3}H_{4}^{\alpha}\right)\left(H_{1}^{\beta}-H_{2}^{-1}H_{3}H_{4}^{\beta}\right)},\\
\mathcal{F} & = & -H_{5}+H_{2}^{-1}\Omega_{\alpha\beta}H_{4}^{\alpha}H_{4}^{\beta}.\nonumber 
\end{eqnarray}
The one-forms are decomposed as
\begin{eqnarray}
\beta & = & \beta_{0}\left(d\psi+\chi_{a}dx^{a}\right)+\beta_{a}dx^{a},\nonumber \\
\omega & = & \omega_{0}\left(d\psi+\chi_{a}dx^{a}\right)+\omega_{a}dx^{a},
\end{eqnarray}
with the coefficients $\beta_{0},$ $\beta_{a},$ $\omega_{0}$ and
$\omega_{a}$ given by
\begin{eqnarray}
\beta_{0} & = & H_{2}^{-1}H_{3},\nonumber \\
\vec{\nabla}\times\vec{\beta} & = & -\vec{\nabla}H_{3}\ ,\nonumber \\
\omega_{0} & = & H_{6}-H_{2}^{-2}H_{3}\Omega_{\alpha\beta}H_{4}^{\alpha}H_{4}^{\beta}+H_{2}^{-1}\Omega_{\alpha\beta}H_{1}^{\alpha}H_{4}^{\beta}+\frac{1}{2}H_{2}^{-1}H_{3}H_{5}\ ,\nonumber \\
\vec{\nabla}\times\vec{\omega} & = & \Omega_{\alpha\beta}\left(H_{4}^{\alpha}\vec{\nabla}H_{1}^{\beta}-H_{1}^{\alpha}\vec{\nabla}H_{4}^{\beta}\right)+H_{2}\vec{\nabla}H_{6}-H_{6}\vec{\nabla}H_{2}+\frac{1}{2}H_{3}\vec{\nabla}H_{5}-\frac{1}{2}H_{5}\vec{\nabla}H_{3}\ .\nonumber \\
\label{curl omega}
\end{eqnarray}
The three-forms are equal to
\begin{equation}
\hat{G}^{\alpha}=-He^{+}\wedge e^{-}\wedge\tilde{d}\left(H^{-1}j^{\alpha}\right)+j^{\alpha}H^{-1}e^{+}\wedge\tilde{d}\beta+e^{-}\wedge\left[-j^{\alpha}\left(\tilde{d}\omega\right)^{-}+H_{\mathrm{SD}}^{\alpha}\right]+\star_{4}\tilde{d}\left(j^{\alpha}H\right),
\end{equation}
where
\begin{eqnarray}
e^{-} & = & H^{-1}\left(du+\beta\right),\nonumber \\
e^{+} & = & dv+\omega+\frac{1}{2}\mathcal{F}He^{-},\\
H_{\mathrm{SD}}^{\alpha} & = & -D_{b}^{\alpha}\left(d\psi+\chi_{a}dx^{a}\right)\wedge dx^{b}-\frac{1}{2}H_{2}D_{b}^{\alpha}\epsilon_{\ cd}^{b}dx^{c}\wedge dx^{d},\nonumber \\
\vec{D}^{\alpha} & = & H\vec{\nabla}\left(H_{2}^{-1}H_{4}^{\alpha}\right)-j^{\alpha}Hj_{\beta}\vec{\nabla}\left(H_{2}^{-1}H_{4}^{\beta}\right).\nonumber 
\end{eqnarray}
Lastly, the scalars are given by
\begin{equation}
j^{\alpha}=\frac{H_{1}^{\alpha}-H_{2}^{-1}H_{3}H_{4}^{\alpha}}{H}\ .
\end{equation}
When the harmonic functions are taken of the form (\ref{form harmonic functions}),
one has to impose the bubble equations (\ref{Bubble equations}) in order to avoid Dirac string-like singularities.

\subsection{Black strings and other objects with a horizon\label{subsec:Black-string-solutions}}

In \cite{Akyol:2011mh} it is shown that in supergravity coupled
to tensor multiplets, near horizon geometries of black objects are
locally either $\mathbb{R}^{1,1}\times T^{4},$ $\mathbb{R}^{1,1}\times K_{3}$
or $AdS_{3}\times S^{3}.$ In this section we will consider some examples
of the latter local geometry which can correspond to a black string or the uplift of a black ring or black lens. 

\paragraph{Black string.}

When taking a solution of section \ref{subsec:Gibbons-Hawking-base-space},
compactifying the $u-$direction on a circle, taking the harmonic
functions of the form (\ref{form harmonic functions}) with only one
center $\vec{x}_{1}=0$ and requiring the metric to asymptote to $\mathbb{R}\times S^{1}\times\mathbb{R}^{4}/\mathbb{Z}_{m},$
we find a generalization of the single black string solution in \cite{Crichigno:2016lac}. Perhaps the most interesting case is $m=1$, which at infinity corresponds to a black string wrapped around a circle times a flat 5d Minkowski spacetime. For $m\neq 1$ one gets ALE spaces.

The string is wound around the $u-$direction and becomes a black
hole after reduction on the $u-$circle. In appendix \ref{sec:Single-string-solution}
we derive that to get the right asymptotics for the metric, we need
\begin{equation}
\Gamma_{\infty}=\left(\mu_{\infty}^{\alpha},0,0,0,-1,\frac{1}{2}\frac{q}{m}-\frac{1}{m}\Omega_{\alpha\beta}\mu_{\infty}^{\alpha}p^{\beta}\right),
\end{equation}
with 
\begin{equation}
\Omega_{\alpha\beta}\mu_{\infty}^{\alpha}\mu_{\infty}^{\beta}=1.
\end{equation}
Using spherical coordinates $r$, $\theta$, $\phi$ for the $\mathbb{R}^{3}$ part, the metric
of the solution is then given by
\begin{eqnarray}
ds_{6}^{2} & = & 2\left(1+2\frac{\Omega_{\alpha\beta}\mu_{\infty}^{\alpha}\tilde{Q}^{\beta}}{4\sqrt{2}r}+\frac{\Omega_{\alpha\beta}\tilde{Q}^{\alpha}\tilde{Q}^{\beta}}{32r^{2}}\right)^{-1/2}du'\nonumber \\
 &  & \times\left[dv+\frac{J_{\psi}}{8r}\left(d\psi+m\cos(\theta)d\phi\right)+\frac{1}{2}\left(1+\frac{Q}{4r}\right)du'\right]\label{solution asymptotically flat}\\
 &  & +\left(1+2\frac{\Omega_{\alpha\beta}\mu_{\infty}^{\alpha}\tilde{Q}^{\beta}}{4\sqrt{2}r}+\frac{\Omega_{\alpha\beta}\tilde{Q}^{\alpha}\tilde{Q}^{\beta}}{32r^{2}}\right)^{1/2}\left[\frac{r}{m}\left(d\psi+m\cos(\theta)d\phi\right)^{2}+\frac{m}{r}dr^{2}+mrd\Omega_{2}^{2}\right],\nonumber 
\end{eqnarray}
where we defined $u'=u+\frac{q}{m}\psi.$ To make this transformation
well-defined, we have to impose 
\begin{equation}
\frac{4\pi q}{lm}\in\mathbb{Z},
\end{equation}
where $l$ is the length of the $u-$circle. We also defined

\begin{eqnarray}
\tilde{Q}^{\alpha} & \equiv & 4\sqrt{2}\left(\mu^{\alpha}-\frac{1}{m}qp^{\alpha}\right),\nonumber \\
Q & \equiv & 4\left(-n+\frac{1}{m}\Omega_{\alpha\beta}p^{\alpha}p^{\beta}\right),\label{constants single black hole solutions}\\
J_{\psi} & \equiv & 8\left(j-\frac{1}{m^{2}}q\Omega_{\alpha\beta}p^{\alpha}p^{\beta}+\frac{1}{m}\left(\Omega_{\alpha\beta}\mu^{\alpha}p^{\beta}+\frac{1}{2}qn\right)\right).\nonumber 
\end{eqnarray}
The near horizon geometry of this solution is a direct product of
an extremal BTZ black hole and a round $S^{3}/\mathbb{Z}_{m}.$ The
entropy of the black string is given by 
\begin{equation}
S=\frac{A}{4G_{\mathrm{N}}^{(6)}}=2\pi\sqrt{\frac{1}{2}mQ\Omega_{\alpha\beta}\tilde{Q}^{\alpha}\tilde{Q}^{\beta}-J_{\psi}^{2}},\label{Entropy macroscopic}
\end{equation}
where we used conventions in which $G_{\mathrm{N}}^{(6)}=\frac{l\pi}{4}.$

These black holes can be embedded in F-theory and we consider the case $m=1$ for simplicity. We take
an F-theory background $\mathbb{R}\times S^{1}\times\mathbb{R}^{4}\times CY_{3},$
where $CY_{3}$ is a smooth elliptically fibered Calabi-Yau three-fold
$\pi:\ CY_{3}\rightarrow B_{2}.$ The solution corresponds to a D3-brane
wrapped on $S^{1}\times C,$ where $C\subset B_{2}$ is a curve. We
have the set of vertical divisors $D_{\alpha}\equiv\pi^{-1}\left(D_{\alpha}^{b}\right)$,
where $D_{\alpha}^{b}$ are divisors of $B_{2}$ that are chosen such
that 
\begin{equation}
\Omega_{\alpha\beta}=\int_{B_{2}}\omega_{\alpha}\wedge\omega_{\beta}
\end{equation}
for $\omega_{\alpha}$ the two-form classes Poincare dual to $D_{\alpha}$.
We can then write $C=q^{\alpha}\omega_{\alpha}$ for the form Poincare
dual to the curve $C.$ The strings that one gets after compactification
on $CY_{3}$ carry $n$ units of momentum along the circle. There
is also an $SO(4)\equiv SU(2)_{L}\times SU(2)_{R}$ symmetry from
rotations in the $\mathbb{R}^{4}$ plane transverse to the D3-brane.
The entropy corresponding to a single string to leading order in the
large charge limit is given by \cite{Haghighat:2015ega}
\begin{equation}
S=2\pi\sqrt{\frac{1}{2}n\Omega_{\alpha\beta}q^{\alpha}q^{\beta}-J^{2}},\label{microscopic entropy stefan vafa}
\end{equation}
where $J$ is the eigenvalue corresponding to the $U(1)_{L} \subset SU(2)_{L}$ symmetry in the convention that it is half-integer valued. The microscopic formula (\ref{microscopic entropy stefan vafa}) is
only valid in the limit where $n$ is much larger than the other charges.
Comparison with (\ref{Entropy macroscopic}) leads to the identification
$Q=n,$ $\tilde{Q}^{\alpha}=q^{\alpha}$ and $J_{\psi}=J$ which explains
the normalizations in (\ref{constants single black hole solutions}). 

One can also construct black string solutions with a Taub-NUT base space and asymptotics $\mathbb{R} \times S^{1} \times S^{1} \times \mathbb{R}^{3}$. Although the full solution will be very different from (\ref{solution asymptotically flat}), the near horizon geometry will be the same. We can then compare (\ref{Entropy macroscopic}) for $m\neq1$ with the leading order contribution of the entropy calculated in the microscopic setting corresponding to the Taub-NUT solution. This setting is an F-theory background $\mathbb{R} \times S^{1} \times TN_{m} \times CY_{3},$ where $TN_{m}$ is a Taub-NUT spacetime with NUT charge $m.$ The D3-brane is still wrapped on $S^{1}\times C$ and given $n$ units
of momentum along the circle after compactification on $CY_{3}.$
The Taub-NUT breaks the $SO(4)$ symmetry to $U(1)_{L}/\mathbb{Z}_{m}\times SU(2)_{R}.$
In \cite{Bena:2006qm} the entropy for this setting is calculated
in the dual picture that one gets starting from type IIB, T-dualizing
along the NUT-circle and then lifting it to M-theory. The M-theory
picture is then given by an M5-brane wrapped around $S^{1}\times\left(mB_{2}+\hat{C}\right),$
where $\hat{C}=\pi^{-1}(C)$. The entropy is calculated using
the MSW formula \cite{Maldacena:1997de} and is to leading order equal
to
\begin{equation}
S=2\pi\sqrt{\frac{1}{2}mn\Omega_{\alpha\beta}q^{\alpha}q^{\beta}-J^{2}},\label{Microscopic entropy Taub-NUT}
\end{equation}
where $J$ is the eigenvalue corresponding to the $U(1)_{L}/\mathbb{Z}_{m}$
symmetry along the NUT-circle. The microscopic formula (\ref{Microscopic entropy Taub-NUT})
is also only valid under certain conditions. Besides the Cardy limit
in which $n$ has to be much larger than the other charges, we also
have that $q^{\alpha}\gg mc^{\alpha},$ where $c^{\alpha}$ comes
from the expansion of the first Chern class of the base space: $c_{1}(B_{2})=c^{\alpha}\omega_{\alpha}.$
The latter condition is needed to make the divisor $mB_{2}+\hat{C}$
very ample. Comparing  (\ref{Entropy macroscopic}) with  (\ref{Microscopic entropy Taub-NUT}) we find that $m$ has to be
an integer and that we can identify $Q=n,$ $\tilde{Q}^{\alpha}=q^{\alpha}$
and $J_{\psi}=J$ in the limits where (\ref{Microscopic entropy Taub-NUT}) is valid. To fully compare this microscopic setting with a macroscopic solution, we of course have to construct the solution with Taub-NUT base space $\mathcal{B}$, but we will leave this for future work \cite{Grimm:2018weo}. 

\paragraph{Uplift black ring.}

A five-dimensional black ring solution  \cite{Emparan:2001wn,Elvang:2004rt,Gauntlett:2004qy} is asymptotically flat, has a regular horizon with topology $S^{1}\times S^{2}$ and near horizon geometry $AdS_{3}\times S^{2}$.
We will show that the 6d uplift has horizon $S^{1}\times S^{3}$
and near horizon geometry $AdS_{3}\times S^{3},$ and is thus consistent with the classification we stated at the beginning of this section.
A more general discussion of uplifts of black rings in connection to supertubes was given in \cite{Elvang:2004ds}. Essentially, our discussion below is a particular and simple case of theirs, so we will be rather brief here and only focus on the near horizon geometry.

To be specific, we take the black ring solution from \cite{Elvang:2004rt}
 written in certain coordinates $v,$ $r,$ $\psi',$
$\phi',$ $\theta$ and $\chi$ and in which the near horizon limit
is taken by redefining $r=\epsilon L\tilde{r}/R,$ $v=\tilde{v}/\epsilon$
(where $L$ and $R$ are certain constants) and sending $\epsilon\rightarrow0.$
In this limit, the metric becomes 
\begin{equation}
ds_{5}^{2}=2d\tilde{v}d\tilde{r}+\frac{4L}{q}\tilde{r}d\tilde{v}d\psi'+L^{2}d\psi'^{2}+\frac{q^{2}}{4}\left(d\theta^{2}+\sin^{2}(\theta)d\chi^{2}\right),
\end{equation}
where $q$ is another constant. In the same limit, the vector field
(in the conventions of section \ref{subsec:Reduction-to-five-dimensions})
is given by 
\begin{eqnarray}
A & = & \frac{1}{2q}\left[3Q-q^{2}\right]\frac{C_{1}}{r}dr-\frac{q}{2}\cos(\theta)d\chi,
\end{eqnarray}
where $Q,$ $C_{1}$ are other constants and where we have added the
exact form 
\begin{equation}
\frac{1}{2q}\left(3Q-3q^{2}\right)d\psi'+\frac{q}{2}d\phi'+\frac{q}{2}d\psi'
\end{equation}
to the expression in \cite{Elvang:2004rt}. From section \ref{subsec:Reduction-to-five-dimensions}
we see that the metric of the six-dimensional uplift of a solution in the timelike class of minimal five-dimensional supergravity is given by 
\begin{equation}
ds_{6}^{2}=\left(du+A\right)^{2}+ds_{5}^{2}\ .
\end{equation}
Redefining $du=-\frac{q}{2}du'-\frac{1}{2q}\left[3Q-q^{2}\right]\frac{C_{1}}{r}dr$
(this can be done in the full solution), the metric becomes
\begin{equation}
ds_{6}^{2}=\frac{q^{2}}{4}\left(du'+\cos(\theta)d\chi\right)^{2}+2d\tilde{v}d\tilde{r}+\frac{4L}{q}\tilde{r}d\tilde{v}d\psi'+L^{2}d\psi'^{2}+\frac{q^{2}}{4}\left(d\theta^{2}+\sin^{2}(\theta)d\chi^{2}\right).
\end{equation}
The $u',$ $\theta$ and $\chi$ part form the round metric
on $S^{3},$ where we need $0\leq u'<4\pi$ to make it regular. The
near horizon geometry of the uplift of the black ring is thus $AdS_{3}\times S^{3}.$

\paragraph*{Black lens.}

In minimal five-dimensional supergravity one also has solutions
that have a horizon with lens space topology $L(m,1)=S^{3}/\mathbb{Z}_{m}$
and are asymptotically flat \cite{Kunduri:2014kja,Tomizawa:2016kjh}.
In section \ref{subsec:Reduction-to-five-dimensions} is described how
such solutions can be uplifted to six dimensions. They will fall in
the class with a Gibbons-Hawking base space and have harmonic functions
of the form (\ref{form harmonic functions}) with $m$ centers. Their near horizon geometry will locally be given by $AdS_{3} \times S^{3}/\mathbb{Z}_{m}$ and their asymptotics will be $\mathbb{R}\times S^{1}\times\mathbb{R}^{4}.$
Note that solutions with the same near horizon geometry but different
asymptotics are given by the previously described black string with
asymptotics $\mathbb{R}\times S^{1}\times\mathbb{R}^{4}/\mathbb{Z}_{m}$
and by a black string with a Taub-NUT space as base space which will
have asymptotics $\mathbb{R}\times S^{1}\times S^{1}\times\mathbb{R}^{3}.$ 

\subsection{pp-waves}

A pp-wave is characterized by the existence of a covariant constant
null vector field. This vector field is necessarily a Killing vector
field. Requiring the null Killing vector field $\partial_{v}$ of
the general solution to be covariantly constant implies that 
\begin{equation}
d\left[H^{-1}\left(du+\beta\right)\right]=0,
\end{equation}
which is equivalent to $\mathcal{D}\beta=0$ and $H^{-1}\left(\mathcal{D}H\right)=-\dot{\beta}$. 

A first class of pp-waves is given by non-twisting solutions of section
\ref{subsec:Non-twisting-solutions} with
$H=H(u)$. It follows from the construction of the coordinates in section \ref{subsec:Introduction-of-coordinates} that in this case we may choose $H=1$ by redefining $u$ in \eqref{XandY}, such
that the solution becomes
\begin{eqnarray}
ds_{6}^{2} & = & 2du\left(dv+\omega+\frac{1}{2}\mathcal{F}du\right)+ds_{4}^{2},\nonumber \\
\hat{G}^{\alpha} & = & -e^{+}\wedge du\wedge\tilde{d}j^{\alpha}+du\wedge\left[j^{\alpha}\psi-j^{\alpha}\left(\tilde{d}\omega\right)^{-}+H_{\mathrm{SD}}^{\alpha}\right]+\star_{4}\tilde{d}j^{\alpha}.
\end{eqnarray}
The flat base space solution derived in section \ref{subsec:Non-twisting-solutions}
is an example of a pp-wave when we take the functions $P^{\alpha}$
and $Q^{\alpha}$ such that $P_{\alpha}P^{\alpha}=1,$ $P_{\alpha}Q^{\alpha}=0$
and $Q_{\alpha}Q^{\alpha}=0.$ This is only possible when $Q^{\alpha}=0.$
Even with all these extra conditions, the tensor branch of the theory
provides a generalization of the solution in \cite{Gutowski:2003rg}
since in general the scalars are still $u-$dependent and the two-forms
$H_{\mathrm{SD}}^{\alpha}$ are non-vanishing. To simplify a bit more
we choose $\eta=1,$ $W_{1}=0$ and take $C_{3}$ such that $\Omega_{\alpha\beta}C_{3}^{\alpha}P^{\beta}=2.$
Transforming to Cartesian coordinates (see \cite{Gauntlett:2002nw})
we find that 
\begin{equation}
\omega=\frac{1}{4}\Omega_{\alpha\beta}C_{3}^{\alpha}P^{\beta}r^{2}\sigma^{3}_{R}=x^{1}dx^{2}-x^{2}dx^{1}+x^{3}dx^{4}-x^{4}dx^{3}.
\end{equation}
For this solution also $\left(\tilde{d}\omega\right)^{-}=0.$ Performing
now the coordinate transformation 
\begin{eqnarray}
x^{1} & = & \sin(u)y^{1}-\cos(u)y^{2},\nonumber \\
x^{2} & = & \cos(u)y^{1}+\sin(u)y^{2},\\
x^{3} & = & \cos(u)y^{3}+\sin(u)y^{4},\nonumber \\
x^{4} & = & -\sin(u)y^{3}+\cos(u)y^{4},\nonumber 
\end{eqnarray}
we obtain the plane wave solution
\begin{eqnarray}
ds_{6}^{2} & = & 2dudv+\left(\mathcal{F}-\delta_{mn}y^{m}y^{n}\right)du^{2}+\delta_{mn}dy^{m}dy^{n},\nonumber \\
\hat{G}^{\alpha} & = & \left(C_{3}^{\alpha}-2P^{\alpha}\right)du\wedge\left[dy^{1}\wedge dy^{2}+dy^{3}\wedge dy^{4}\right],\label{eq:non-twisting solution and pp-wave}\\
j^{\alpha} & = & P^{\alpha}(u).\nonumber 
\end{eqnarray}

A second class of pp-waves are the solutions in which $\hat{G}^{\alpha}$
vanish. A subset of these solutions is given by the vacuum solutions
in which also the (physical) scalars vanish. From (\ref{threeforms using coordinates})
we find that $\hat{G}^{\alpha}=0$ is equivalent to
\begin{eqnarray}
H^{-1}\mathcal{D}H & = & -\dot{\beta},\nonumber \\
\mathcal{D}\beta & = & 0,\nonumber \\
\mathcal{D}j^{\alpha} & = & 0,\label{eq:conditions vanishing threeforms}\\
H\psi & = & \left(\mathcal{D}\omega\right)^{-},\nonumber \\
H_{\mathrm{SD}}^{\alpha} & = & 0.\nonumber 
\end{eqnarray}
The first and second conditions define a pp-wave. The set of equations
(\ref{eq:conditions vanishing threeforms}) will be hard to solve
without extra assumptions. Of course, one can again look at the subclasses
of non-twisting and $u-$independent solutions. As an example of a
non-twisting solution that falls in this class, we can take (\ref{eq:non-twisting solution and pp-wave})
with $C_{3}^{\alpha}=2P^{\alpha}.$ In this case the solution simplifies
to 
\begin{eqnarray}
ds_{6}^{2} & = & 2dudv+\left(C_{5}(u)-\frac{1}{2}C_{4}(u)\frac{1}{r^{2}}+\frac{1}{4}\Omega_{\alpha\beta}\dot{P}^{\alpha}\dot{P}^{\beta}r^{2}\right)du^{2}+\delta_{mn}dy^{m}dy^{n},\nonumber \\
\hat{G}^{\alpha} & = & 0,\label{pp wave with vanishing three-form}\\
j^{\alpha} & = & P^{\alpha}(u),\nonumber 
\end{eqnarray}
where we still have the condition $P_{\alpha}P^{\alpha}=1.$ 

One last example we consider is not a proper pp-wave, but it is a
black string with traveling waves that carry momentum along the string
\cite{Horowitz:1996th,Horowitz:1996cj}. This solution falls into the non-twisting class with
flat base space (section \ref{subsec:Non-twisting-solutions}). Taking $W_{1}=C_{3}^{\alpha}=C_{5}=0$ and $P^{\alpha}$
and $Q^{\alpha}$ constant such that $\Omega_{\alpha\beta}P^{\alpha}P^{\beta}=1,$
we find the solution
\begin{eqnarray}
ds_{6}^{2} & = & 2H^{-1}du\left(dv-\frac{1}{4}C_{4}(u)\frac{1}{r^{2}}du\right)+Hds_{4}^{2}\ ,\nonumber \\
\hat{G}^{\alpha} & = & -dv\wedge du\wedge\tilde{d}\left(H^{-1}j^{\alpha}\right)+\star_{4}\tilde{d}\left(j^{\alpha}H\right)\ ,\\
j^{\alpha}H & = & P^{\alpha}+\frac{Q^{\alpha}}{r^{2}}\ .\nonumber 
\end{eqnarray}
In the limit $r\rightarrow\infty$ this solution is the same as (\ref{pp wave with vanishing three-form})
with $C_{5}=\dot{P}^{\alpha}=0,$ but note that the full solution
is very different, mainly because we now have non-vanishing three-forms
to support the black string and also the scalars depend on the base
space instead of on $u.$ For a further discussion of this kind of geometry,
see e.g. the original references \cite{Horowitz:1996th,Horowitz:1996cj}. The horizon of these solutions become singular however, as discussed e.g. in \cite{Kaloper:1996hr,Horowitz:1997si,Ross:1997pd}. Perhaps a more interesting class of solutions are the traveling wave deformations of smooth horizonless solutions, such as discussed e.g. in \cite{Lunin:2012gp}. It could be interesting to extend the analysis of \cite{Lunin:2012gp} to the present setup where more tensor multiplets are present.

\section{Attractor mechanism \label{sec:Attractor-Mechanism}}

In this section we study the attractor mechanism \cite{Ferrara:1995ih,Ferrara:1996dd}
in six-dimensional $(1,0)$ supergravity coupled to tensor
multiplets. We first repeat the near horizon analysis \cite{Andrianopoli:2007kz,Ferrara:2008xz}
to show that the scalars near the horizon can be expressed in terms
of the charges of the black object. After that we derive a ``flow''
equation for $u-$independent solutions which in certain simplifying
cases explains this attractor mechanism from the full geometry perspective.
Our version of the attractor flow is consistent with the five-dimensional
flow equation in \cite{Kraus:2005gh}. A general proof of the attractor mechanism for single, charged, static, flat $p-$brane solutions in $d$ dimensions is given in \cite{deAntonioMartin:2012bi} . Some of the solutions we consider will also be of this type, but not all of them.

\subsection{Near horizon analysis}

We consider the near horizon geometries of black objects which are
locally $AdS_{3}\times S^{3}$ (so we consider one of the three possible cases). In \cite{Akyol:2011mh} it is shown
that in this geometry the tensor multiplet scalars $j^{\alpha}$ are
constants and the tensors $H^{M}$ of the tensor multiplets vanish.
Integrating over the spherical part of the solution (e.g. in (\ref{solution asymptotically flat})
this part is parametrized by $\psi,$ $\phi$ and $\theta$) implies that the charges that correspond
to $\hat{G}^{\alpha}=j^{\alpha}H$ are equal to 
\begin{equation}
\tilde{Q}^{\alpha}=j^{\alpha}k.
\end{equation}
Using $j_{\alpha}j^{\alpha}=1,$ we find that 
\begin{equation}
k=\sqrt{\Omega_{\alpha\beta}\tilde{Q}^{\alpha}\tilde{Q}^{\beta}}
\end{equation}
and 
\begin{equation}
j^{\alpha}=\frac{\tilde{Q}^{\alpha}}{\sqrt{\Omega_{\beta\gamma}\tilde{Q}^{\beta}\tilde{Q}^{\gamma}}}.\label{scalars near-horizon}
\end{equation}
Hence in the near horizon geometry the scalars take a value completely
expressed in terms of the charges related to the three-forms. 

\subsection{Flow equation }

It would be nice to be able to see the scalar values arise from the
flow of a central charge as one usual can (e.g. \cite{Kraus:2005gh}).\textbf{
}We will derive this ``flow'' equation for $u-$independent solutions.
The general flow is complicated, but we consider a class of solutions
where it simplifies. To derive the flow equation, we need two ingredients:
the Bianchi identity and supersymmetry. The part of the three-forms
with three legs on the base space is generally what corresponds to the charges,
hence we will derive an equation for the scalars and $\hat{G}_{ijk}^{\alpha}$ (note however, that in the near horizon geometry of the uplift of the black ring, the three-sphere is given by the $u-$circle fibered over an $S^{2}$ in the base space).
Using supersymmetry and $u-$independence (\ref{threeforms u-independent solutions}),
but not specifying the part $\hat{G}_{ijk}^{\alpha}$ we can write
\begin{eqnarray}
\hat{G}^{\alpha} & = & -He^{+}\wedge e^{-}\wedge\tilde{d}\left(H^{-1}j^{\alpha}\right)+j^{\alpha}H^{-1}e^{+}\wedge\tilde{d}\beta+e^{-}\wedge\left[-j^{\alpha}\left(\tilde{d}\omega\right)^{-}+H_{\mathrm{SD}}^{\alpha}\right]\nonumber \\
 &  & +\frac{1}{6}\hat{G}_{ijk}^{\alpha}\tilde{e}^{i}\wedge\tilde{e}^{j}\wedge\tilde{e}^{k}.\label{General three-form}
\end{eqnarray}
We now first consider the Bianchi identity and after that use the
tensorini equation to finish the derivation of the flow equation.
Since we are only interested in $\hat{G}_{ijk}^{\alpha}$ we will,
after applying the exterior derivative on (\ref{General three-form}),
only consider the part with four legs on $\mathcal{B}:$ 

\begin{eqnarray}
d\hat{G}^{\alpha} & \rightarrow & j^{\alpha}H^{-1}de^{+}\wedge\tilde{d}\beta+de^{-}\wedge\left[-j^{\alpha}\left(\tilde{d}\omega\right)^{-}+H_{\mathrm{SD}}^{\alpha}\right]+d\left(\frac{1}{6}\hat{G}_{ijk}^{\alpha}\tilde{e}^{i}\wedge\tilde{e}^{j}\wedge\tilde{e}^{k}\right).\nonumber \\
\label{eq:bianchi}
\end{eqnarray}
Calculating the last term in (\ref{eq:bianchi}) yields
\begin{eqnarray}
d\left(\hat{G}_{ijk}^{\alpha}\tilde{e}^{i}\wedge\tilde{e}^{j}\wedge\tilde{e}^{k}\right) & = & \tilde{\nabla}_{l}\left(\hat{G}_{ijk}^{\alpha}\right)\tilde{e}^{l}\wedge\tilde{e}^{i}\wedge\tilde{e}^{j}\wedge\tilde{e}^{k}.\label{2nd part bianchi}
\end{eqnarray}
Using that
\begin{equation}
de^{-}  \rightarrow  H^{-1}\tilde{d}\beta\ ,\qquad
de^{+}  \rightarrow  \tilde{d}\omega+\frac{1}{2}\mathcal{F}\tilde{d}\beta\ ,
\end{equation}
we can finish the calculation of (\ref{eq:bianchi}):
\begin{eqnarray}
d\hat{G}^{\alpha} & \rightarrow & j^{\alpha}H^{-1}\left(\tilde{d}\omega+\frac{1}{2}\mathcal{F}\tilde{d}\beta\right)\wedge\tilde{d}\beta+H^{-1}\tilde{d}\beta\wedge\left[-j^{\alpha}\left(\tilde{d}\omega\right)^{-}+H_{\mathrm{SD}}^{\alpha}\right]\nonumber \\
 &  & +d\left(\frac{1}{6}\hat{G}_{ijk}^{\alpha}\tilde{e}^{i}\wedge\tilde{e}^{j}\wedge\tilde{e}^{k}\right)\nonumber \\
 & = & H^{-1}\left[j^{\alpha}\left(\tilde{d}\omega\right)^{+}+\frac{1}{2}j^{\alpha}\mathcal{F}\tilde{d}\beta+H_{\mathrm{SD}}^{\alpha}\right]\wedge\tilde{d}\beta+d\left(\frac{1}{6}\hat{G}_{ijk}^{\alpha}\tilde{e}^{i}\wedge\tilde{e}^{j}\wedge\tilde{e}^{k}\right).
\end{eqnarray}
With (\ref{2nd part bianchi}) this can be rewritten as
\begin{equation}
d\hat{G}^{\alpha}\rightarrow\left[\frac{1}{4}\left(\mathcal{G}^{+\alpha}\right)_{li}\left(\tilde{d}\beta\right)_{jk}+\frac{1}{6}\tilde{\nabla}_{l}\left(\hat{G}_{ijk}^{\alpha}\right)\right]\tilde{e}^{l}\wedge\tilde{e}^{i}\wedge\tilde{e}^{j}\wedge\tilde{e}^{k}.
\end{equation}
The Bianchi identity implies that
\begin{eqnarray}
\frac{1}{6}\tilde{\nabla}_{l}\left(\hat{G}_{ijk}^{\alpha}\right) & \epsilon^{lijk} & =-\frac{1}{4}\left(\mathcal{G}^{+\alpha}\right)_{li}\left(\tilde{d}\beta\right)_{jk}\epsilon^{lijk}.\label{bianchi identity result}
\end{eqnarray}
From supersymmetry (\ref{threeforms u-independent solutions}) it
follows that
\begin{equation}
\hat{G}_{ijk}^{\alpha}=\tilde{d}\left(j^{\alpha}H\right)_{l}\epsilon_{\ ijk}^{l}
\end{equation}
such that
\begin{equation}
\partial_{l}\left(j^{\alpha}\right)\Omega_{\alpha\beta}\hat{G}_{ijk}^{\beta}\epsilon^{lijk}=-6Hg_{\alpha\beta}\partial_{l}\left(j^{\alpha}\right)\partial^{l}\left(j^{\beta}\right).
\end{equation}
Combining this with (\ref{bianchi identity result}) results in 
\begin{eqnarray}
\frac{1}{6}\tilde{\nabla}_{l}\left(\Omega_{\alpha\beta}j^{\alpha}\hat{G}_{ijk}^{\beta}\right)\epsilon^{lijk} & = & -Hg_{\alpha\beta}\partial_{l}\left(j^{\alpha}\right)\partial^{l}\left(j^{\beta}\right)-j_{\alpha}\mathcal{G}^{+\alpha}\cdot\tilde{d}\beta.\label{Flow equation u-indep}
\end{eqnarray}
This is the flow equation for $u-$independent solutions. In principle
one can also do this derivation for the most general solution, but
the resulting equation does not put a strong constraint on a flow. Even for the most general $u-$independent
solution the meaning of (\ref{Flow equation u-indep}) is not very
clear. However, when also either $\tilde{d}\beta$ or $j_{\alpha}\mathcal{G}^{+\alpha}$
vanish, the right-hand side of (\ref{Flow equation u-indep}) is non-positive
since $g_{\alpha\beta}$ is positive definite. This implies a monotonicity
property of the quantity $\frac{1}{6}\Omega_{\alpha\beta}j^{\alpha}\hat{G}_{ijk}^{\beta}\epsilon^{lijk}$. When one derives the flow equation for the most general solution there might be other special cases in which the equation implies a monotonicity property of a quantity. One can also derive an equation for other components of $\hat{G}^{\alpha}$ and examine what this equation would imply for black ring solutions. As a check, we show in appendix \ref{Sec. reduction flow equation to five dim}
that when one performs a Kaluza-Klein reduction along the $u-$circle
(section \ref{subsec:Reduction-to-five-dimensions}), the flow equation
(\ref{Flow equation u-indep}) reduces to the five-dimensional flow
equation derived in \cite{Kraus:2005gh}.

If we let $V\subset\mathcal{B}$ we can define the charges
\begin{eqnarray}
\tilde{Q}^{\alpha} & = & \frac{1}{12\sqrt{2}\pi^{2}}\int_{\partial V}dS\,\hat{G}_{ijk}^{\alpha}n_{l}\epsilon^{lijk}\ ,\label{charge general solutions gh}
\end{eqnarray}
where $n$ is a unit vector perpendicular to $\partial V$ and pointing
outward. The quantity (\ref{charge general solutions gh}) is for
the black string solution in (\ref{solution asymptotically flat}) equal to
the charge defined in (\ref{constants single black hole solutions}).
We can also introduce the central charge
\begin{equation}
Z(V)\equiv\frac{1}{12\sqrt{2}\pi^{2}}\int_{\partial V}dS\,\Omega_{\alpha\beta}j^{\alpha}\hat{G}_{ijk}^{\beta}n_{l}\epsilon^{lijk},\label{electric central charge}
\end{equation}
which, in case the scalars are independent of the region $\partial V,$
reduces to
\begin{equation}
Z(V)=j_{\alpha}\tilde{Q}^{\alpha}.
\end{equation}
This is the central charge that follows from the supersymmetry algebra \cite{Ferrara:1996wv}. When we have regions $V_{2}\subset V_{1},$ we can use (\ref{Flow equation u-indep})
to show that 
\begin{equation}
Z(V_{1})-Z(V_{2})=\frac{1}{2\sqrt{2}\pi^{2}}\int d^{4}x\sqrt{h}\left(-Hg_{\alpha\beta}\partial_{l}\left(j^{\alpha}\right)\partial^{l}\left(j^{\beta}\right)-\frac{1}{2}j_{\alpha}\mathcal{G}^{+\alpha}\cdot\tilde{d}\beta\right).\label{eq:central charge difference}
\end{equation}
In case either $j_{\alpha}\mathcal{G}^{+\alpha}$ or $\tilde{d}\beta$
vanishes, the central charge is monotonically decreasing as we move
outwards. If this is not the case, the flow equation does not provide
a strong constraint on the flow. 

\subsection{One-centered Gibbons-Hawking}

When we take a general solution of section \ref{subsec:Gibbons-Hawking-base-space}
with harmonic functions of the form (\ref{form harmonic functions})
and one center, we actually have a clear radial direction such that
a natural choice of subspaces $V\subset\mathcal{B}$ is $r\leq r_{0}.$
For the general case
\begin{eqnarray}
Z(r_{0}) & = & \frac{1}{12\sqrt{2}\pi^{2}}\int_{\partial V}dS\, \Omega_{\alpha\beta}j^{\alpha} \hat{G}_{ijk}^{\beta}n_{l}\epsilon^{lijk}=\Omega_{\alpha\beta}j^{\alpha}\tilde{Q}^{\beta}
\end{eqnarray}
and from (\ref{eq:central charge difference}) we find that when either
$j_{\alpha}\mathcal{G}^{+\alpha}$ or $\tilde{d}\beta$ vanishes
\begin{eqnarray}
\partial_{r}Z(r) & = & \frac{1}{2\sqrt{2}\pi^{2}}\partial_{r}\int d^{4}x\sqrt{h}\left[-Hg_{\alpha\beta}\partial_{l}\left(j^{\alpha}\right)\partial^{l}\left(j^{\beta}\right)\right]\nonumber \\
 & = & -4\sqrt{2}r^{2}H_{2}Hg_{\alpha\beta}\partial_{r}\left(j^{\alpha}\right)\partial^{r}\left(j^{\beta}\right),
\end{eqnarray}
which is non-positive. Notice that we can write this as
\begin{equation}
r\partial_{r}Z=-4\sqrt{2}r^{3}H_{2}H^{2}\epsilon,\label{near horizon attractor}
\end{equation}
where 
\begin{equation}
\epsilon=\hat{g}^{rr}g_{\alpha\beta}\partial_{r}\left(j^{\alpha}\right)\partial_{r}\left(j^{\beta}\right)
\end{equation}
is the energy density of the scalar fields. Note that near the horizon
$HH_{2}\sim\frac{1}{r^{2}}$ which implies that the proper distance
to the horizon blows up. Together with a finite area of the horizon,
this implies that $\epsilon=0$ because otherwise the energy of the
scalar fields would diverge. Hence from (\ref{near horizon attractor})
we find that at the horizon we get
\begin{equation}
r\partial_{r}Z=0,\label{horizon spaetime form attractor formula}
\end{equation}
which is the spacetime form of the attractor formula. 

For the most general solution with harmonic functions of the form
(\ref{form harmonic functions}), the charge (\ref{charge general solutions gh})
reduces in the near horizon limit $r\rightarrow0$ to 
\begin{equation}
\tilde{Q}^{\alpha}\rightarrow4\sqrt{2}\left(\mu^{\alpha}-\frac{1}{m}qp^{\alpha}\right).
\end{equation}
In the same limit 
\begin{equation}
j^{\alpha}\rightarrow\frac{\mu^{\alpha}-\frac{1}{m}qp^{\alpha}}{\sqrt{\Omega_{\beta\gamma}\left(\mu^{\beta}-\frac{1}{m}qp^{\beta}\right)\left(\mu^{\gamma}-\frac{1}{m}qp^{\gamma}\right)}}=\frac{\tilde{Q}^{\alpha}}{\sqrt{\Omega_{\beta\gamma}\tilde{Q}^{\beta}\tilde{Q}^{\gamma}}}
\end{equation}
which is indeed the value we found in (\ref{scalars near-horizon}).
This is also true for the cases where the flow is more complicated.

\section{Outlook \label{sec:Summary-and-Discussion}}

We derived and analyzed the general local form of supersymmetric solutions of  $(1,0)$
supergravity coupled to tensor multiplets, and studied examples of BPS black strings and pp-waves with non-trivial scalar profiles. 
There are many interesting extensions and generalizations, such as the study of bound states of black strings, and the construction of new microstate geometries and their dual CFT states. It would also be interesting to repeat the general analysis to the case with
hypermultiplets and vector multiplets. 

We solved the equations of motion completely in certain simplifying
cases and studied the attractor flow for $u-$independent solutions. Something that might be interesting as well is to see
if there are also attractor mechanisms for the hyperscalars. For
maximally supersymmetric solutions this is certainly the case \cite{Akyol:2011mh},
but it is not so clear when the solutions have less Killing spinors.

While we have studied to some extent the embedding in type IIB and in F-theory, it would be nice to study better the microscopic analysis of the black string solutions in F-theory. In particular, the near horizon geometry of black strings leads to new dual (0,4) CFTs that are yet to be constructed and analyzed. For the case of minimal supergravity, corresponding to F-theory compactified on a CY$_3$ with base space $\mathbb{P}^2$, this was done in \cite{Haghighat:2015ega}, see also \cite{Vafa:1997gr} for earlier work, and \cite{Lawrie:2016axq,Couzens:2017way,Couzens:2017nnr} for more recent work.

\paragraph*{Acknowledgements.}

We would like to thank Kilian Mayer and Thomas Grimm for useful discussions. We are also grateful to Kilian Mayer for reading and commenting this document. We also thank many of the participants of the "Strings, Geometry and Black Holes" conference (Eurostrings 2018) in London, for useful comments and feedback on the poster presentation of this paper. This work was supported in part by the
D-ITP consortium, a program of the Netherlands Organization for Scientific
Research (NWO) that is funded by the Dutch Ministry of Education,
Culture and Science (OCW), and by the NWO Graduate Programme.

\section*{Appendices}

{\small{}\addcontentsline{toc}{section}{Appendices}}{\small \par}

\addtocontents{toc}{\setcounter{tocdepth}{-1}}

\appendix

\section{Integrability conditions\label{sec:Integrability-conditions}}

In this appendix we derive which equations of motion are implied by
the integrability conditions of the theory. Denote the scalar equations
of motion by $\left(Ej\right)^{M}=0,$ the three-form equations of
motion by $\left(E\hat{G}\right)_{\alpha}^{\mu\nu}=\hat{\nabla}_{\lambda}\left(g_{\alpha\beta}\hat{G}^{\beta\ \lambda\mu\nu}\right)=0$
and the Einstein equation by $E_{\mu\nu}=0.$ Contracting the integrability
of the gravitino Killing spinor equation with $\gamma^{\nu}$ yields\footnote{This condition is derived in the PhD thesis of Mehmet Akyol (Kings
college).}
\begin{equation}
\gamma^{\nu}[\mathcal{D}_{\mu},\mathcal{D}_{\nu}]\epsilon=E_{\mu\nu}\gamma^{\nu}\epsilon+\frac{1}{8}j^{\alpha}\left(E\hat{G}\right)_{\alpha\ \rho\sigma}\hat{g}_{\mu\nu}\gamma^{\nu\rho\sigma}\epsilon-\frac{1}{4}j^{\alpha}\left(E\hat{G}\right)_{\alpha\ \mu\nu}\gamma^{\nu}\epsilon=0.
\end{equation}
Assuming the three-form equations of motion it follows that 
\begin{equation}
E_{\mu\nu}\gamma^{\nu}\epsilon=0.\label{integrability einstein}
\end{equation}
In the null-basis $\gamma_{+}\epsilon=0,$ thus we observe that (\ref{integrability einstein})
implies that 
\begin{equation}
E_{\mu+}=E_{\mu1}=E_{\mu2}=E_{\mu3}=E_{\mu4}=0.
\end{equation}
Hence, only the $E_{--}$ component is not determined by the integrability
conditions.

The integrability of the tensorini Killing spinor equation contracted
with $\gamma^{\mu}$ and expressed in the equations of motion yields\footnote{This condition is derived in the PhD thesis of Mehmet Akyol (Kings
college), although there the wrong scalar equation of motion is used.}
\begin{equation}
\gamma^{\mu}[\mathcal{D}_{\mu},T_{\nu}^{M}\gamma^{\nu}-\frac{1}{12}H_{\nu\rho\sigma}^{M}\gamma^{\nu\rho\sigma}]\epsilon=\left(Ej\right)^{M}\epsilon+\frac{1}{2}x_{\alpha}^{M}\left(E\hat{G}\right)_{\mu\nu}^{\alpha}\gamma^{\mu\nu}\epsilon=0.\label{integrability scalar}
\end{equation}
Assuming the three-form equations of motion, it follows that (\ref{integrability scalar})
is equivalent to the scalar equations of motion.

\section{Spin connection\label{sec:Spin-connection}}

In this section, the components $i,$ $j,...$ will refer to the part
of the six-dimensional vielbein $\hat{e}^{i},$ unless they are components
of base space objects $\tilde{\omega}$ and $\tilde{e}^{i}.$ Using
metric compatibility, so anti-symmetry of the connection 
\begin{eqnarray}
\hat{\omega}_{ij} & = & -\hat{\omega}_{ji},\nonumber \\
\hat{\omega}_{+i} & = & -\hat{\omega}_{i+},\nonumber \\
\hat{\omega}_{-i} & = & -\hat{\omega}_{i-},\\
\hat{\omega}_{++} & = & 0,\nonumber \\
\hat{\omega}_{--} & = & 0,\nonumber \\
\hat{\omega}_{+-} & = & -\hat{\omega}_{-+},\nonumber 
\end{eqnarray}
and vanishing torsion, a straightforward calculation yields that the
spin connection is given by
\begin{eqnarray}
\hat{\omega}_{\ \ i}^{+} & = & \frac{1}{2}\left(\mathcal{D}\omega+\frac{1}{2}\mathcal{F}\mathcal{D}\beta\right)_{ij}\hat{e}^{j}-\frac{1}{2}\partial_{u}\left(Hh_{mn}\right)\tilde{e}_{i}^{m}\tilde{e}_{j}^{n}\hat{e}^{j}\nonumber \\
 &  & -H\left(\dot{\omega}+\frac{1}{2}\mathcal{F}\dot{\beta}-\frac{1}{2}\mathcal{D}\mathcal{F}\right)_{i}e^{-}-\frac{1}{2}\left(H^{-1}\mathcal{D}H+\dot{\beta}\right)_{i}e^{+},\nonumber \\
\hat{\omega}_{\ \ i}^{-} & = & \frac{1}{2}H^{-1}\left(\mathcal{D}\beta\right)_{ij}\hat{e}^{j}-\frac{1}{2}\left(H^{-1}\mathcal{D}H+\dot{\beta}\right)_{i}e^{-},\nonumber \\
\hat{\omega}_{\ \ +}^{+} & = & -\frac{1}{2}\left(H^{-1}\mathcal{D}H+\dot{\beta}\right)_{i}\hat{e}^{i},\\
\hat{\omega}_{\ \ -}^{-} & = & \frac{1}{2}\left(H^{-1}\mathcal{D}H+\dot{\beta}\right)_{i}\hat{e}^{i},\nonumber \\
\hat{\omega}_{\ \,j}^{i} & = & \tilde{\omega}_{\ \,j}^{i}+\frac{1}{2}H^{-1}\left(\mathcal{D}H\right)_{j}\hat{e}^{i}-\frac{1}{2}H^{-1}\left(\mathcal{D}H\right)^{i}\delta_{jk}\hat{e}^{k}+\frac{1}{2}H^{1/2}\left(\beta\wedge\dot{\tilde{e}}^{i}\right)_{kj}\hat{e}^{k}+\frac{1}{2}H^{1/2}\left(\beta\wedge\dot{\tilde{e}}_{k}\right)_{\ \,j}^{i}\hat{e}^{k}\nonumber \\
 &  & +\frac{1}{2}H^{1/2}\left(\beta\wedge\dot{\tilde{e}}_{j}\right)_{\ \,k}^{i}\hat{e}^{k}-\frac{1}{2}H^{-1}\left(\mathcal{D}\beta\right)_{\ \,j}^{i}e^{+}-\frac{1}{2}\left(\mathcal{D}\omega+\frac{1}{2}\mathcal{F}\mathcal{D}\beta\right)_{\ \,j}^{i}e^{-}-H\left(\partial_{u}\tilde{e}_{m}^{[i}\right)\tilde{e}_{j]}^{m}e^{-}.\nonumber 
\end{eqnarray}

\section{$R_{--}$ component of the Ricci tensor \label{sec:-component-of}}

In this section, the components $i,$ $j,...$ will at first refer
to the part of the six-dimensional vielbein $\hat{e}^{i},$ unless
they are components of base space objects $\tilde{\omega}$ and $\tilde{e}^{i}$
or indicated as $\tilde{i}.$ We would like to calculate the $--$
component of the Ricci tensor:
\begin{equation}
R_{--}=R_{\ \ -+-}^{+}+R_{\ \ -i-}^{i}.
\end{equation}
We calculate the curvature two-form via the spin-connection: \textbf{
\begin{eqnarray}
R_{\ \,-}^{i} & = & d\hat{\omega}_{\ \,-}^{i}+\hat{\omega}_{\ \,j}^{i}\wedge\hat{\omega}_{\ \,-}^{j}+\hat{\omega}_{\ \,-}^{i}\wedge\hat{\omega}_{\ \ -}^{-},\nonumber \\
R_{\ \ -}^{+} & = & d\hat{\omega}_{\ \ -}^{+}+\hat{\omega}_{\ \ i}^{+}\wedge\hat{\omega}_{\ \,-}^{i},
\end{eqnarray}
}where we will only keep the $R_{\ \,-j-}^{i}$ and $R_{\ \ -+-}^{+}$
components. A straightforward (but lengthy) calculation yields
\begin{eqnarray}
R_{\ \ -}^{+} & = & 0,\nonumber \\
R_{\ \,-}^{i} & \rightarrow & \Biggl\{ H\partial_{u}\left[\frac{1}{2}\left(\mathcal{D}\omega+\frac{1}{2}\mathcal{F}\mathcal{D}\beta\right)_{\ \,j}^{i}-\frac{1}{2}\partial_{u}\left(Hh_{mn}\right)\tilde{e}^{mi}\tilde{e}_{j}^{n}\right]\nonumber \\
 &  & +\left[\frac{1}{2}\left(\mathcal{D}\omega+\frac{1}{2}\mathcal{F}\mathcal{D}\beta\right)_{\ \,k}^{i}-\frac{1}{2}\partial_{u}\left(Hh_{mn}\right)\tilde{e}^{mi}\tilde{e}_{k}^{n}\right]\left[\frac{1}{2}\dot{H}\delta_{j}^{k}+H\left(\partial_{u}\tilde{e}_{o}^{k}\right)\tilde{e}_{j}^{o}\right]\nonumber \\
 &  & +\frac{1}{2}H^{-1}\left(\partial_{\tilde{j}}H\right)\left(\dot{\omega}+\frac{1}{2}\mathcal{F}\dot{\beta}-\frac{1}{2}\mathcal{D}\mathcal{F}\right)^{\tilde{i}}+\tilde{\nabla}_{\tilde{j}}\left[\left(\dot{\omega}+\frac{1}{2}\mathcal{F}\dot{\beta}-\frac{1}{2}\mathcal{D}\mathcal{F}\right)^{\tilde{i}}\right]\nonumber \\
 &  & -\partial_{u}\left[H\left(\dot{\omega}+\frac{1}{2}\mathcal{F}\dot{\beta}-\frac{1}{2}\mathcal{D}\mathcal{F}\right)^{i}\right]\beta_{j}-\frac{3}{2}H\left(\dot{\omega}+\frac{1}{2}\mathcal{F}\dot{\beta}-\frac{1}{2}\mathcal{D}\mathcal{F}\right)^{i}\left(H^{-1}\mathcal{D}H+\dot{\beta}\right)_{j}\nonumber \\
 &  & -\frac{1}{2}H\left(H^{-1}\mathcal{D}H+\dot{\beta}\right)^{i}\left(\dot{\omega}+\frac{1}{2}\mathcal{F}\dot{\beta}-\frac{1}{2}\mathcal{D}\mathcal{F}\right)_{j}\nonumber \\
 &  & +H\left(\dot{\omega}+\frac{1}{2}\mathcal{F}\dot{\beta}-\frac{1}{2}\mathcal{D}\mathcal{F}\right)^{k}\Biggl[\frac{1}{2}H^{-1}\left(\mathcal{D}H\right)_{k}\delta_{j}^{i}-\frac{1}{2}H^{-1}\left(\mathcal{D}H\right)^{i}\delta_{kj}+\frac{1}{2}H^{1/2}\left(\beta\wedge\dot{\tilde{e}}^{i}\right)_{jk}\nonumber \\
 &  & +\frac{1}{2}H^{1/2}\left(\beta\wedge\dot{\tilde{e}}_{j}\right)_{\ \,k}^{i}+\frac{1}{2}H^{1/2}\left(\beta\wedge\dot{\tilde{e}}_{k}\right)_{\ \,j}^{i}\Biggl]-\left[\frac{1}{2}\left(\mathcal{D}\omega+\frac{1}{2}\mathcal{F}\mathcal{D}\beta\right)_{\ \,k}^{i}+H\left(\partial_{u}e_{m}^{[i}\right)e_{k]}^{m}\right]\nonumber \\
 &  & \times\left[\frac{1}{2}\left(\mathcal{D}\omega+\frac{1}{2}\mathcal{F}\mathcal{D}\beta\right)_{\ \,j}^{k}-\frac{1}{2}\partial_{u}\left(Hh_{mn}\right)\tilde{e}^{mk}\tilde{e}_{j}^{n}\right]\Biggl\}\hat{e}^{j}\wedge e^{-}.
\end{eqnarray}
Taking the $R_{\ \,-i-}^{i}$ components, summing over $i$ and rewriting
everything in components with respect to the vielbein $\tilde{e}^{i}$
yields
\begin{eqnarray}
R_{--} & = & \star_{4}\mathcal{D}\star_{4}\left(\dot{\omega}+\frac{1}{2}\mathcal{F}\dot{\beta}-\frac{1}{2}\mathcal{D}\mathcal{F}\right)-2\left(\dot{\omega}+\frac{1}{2}\mathcal{F}\dot{\beta}-\frac{1}{2}\mathcal{D}\mathcal{F}\right)^{m}\partial_{u}\left(\beta_{m}\right)\nonumber \\
 &  & +\frac{1}{4}H^{-2}\left(\mathcal{D}\omega+\frac{1}{2}\mathcal{F}\mathcal{D}\beta\right)_{ik}\left(\mathcal{D}\omega+\frac{1}{2}\mathcal{F}\mathcal{D}\beta\right)^{ik}\\
 &  & -\frac{1}{2}Hh^{mn}\partial_{u}^{2}\left(Hh_{mn}\right)-\frac{1}{4}\partial_{u}\left(Hh_{mn}\right)\partial_{u}\left(Hh^{mn}\right).\nonumber 
\end{eqnarray}

\section{Single string solution\label{sec:Single-string-solution}}

Taking a single string at the origin, we still have to determine the
one-forms $\vec{\chi},$ $\vec{\beta}$ and $\vec{\omega}$. Solving
the equations for $\vec{\chi},$ $\vec{\beta}$ and $\vec{\omega},$
and assuming the Bubble equations, we find that 
\begin{eqnarray}
\chi_{a}dx^{a} & = & m\cos(\theta)d\phi,\nonumber \\
\beta_{a}dx^{a} & = & -q\cos(\theta)d\phi,\\
\omega_{a}dx^{a} & = & 0.\nonumber 
\end{eqnarray}

\subsection{Asymptotics }

Taking a single string, we require the metric to asymptote to $\mathbb{R}\times S_{u}^{1}\times\mathbb{R}^{4}/\mathbb{Z}_{m}$.
This implies that $m_{\infty}=0,$ the functions $H,\mathcal{F}\rightarrow1$
and the one-forms $\omega,\beta\rightarrow0.$ The limit
\begin{equation}
\lim_{r\rightarrow\infty}\mathcal{F}=-n_{\infty}+\frac{r}{m}\Omega_{\alpha\beta}p_{\infty}^{\alpha}p_{\infty}^{\beta}+2\frac{1}{m}\Omega_{\alpha\beta}p_{\infty}^{\alpha}p^{\beta}=1
\end{equation}
implies that $p_{\infty}^{\alpha}=0$ and $n_{\infty}=-1.$ The limit
of the one-form $\beta\rightarrow0$ implies that $q_{\infty}=0$
such that $\beta\rightarrow\frac{q}{m}d\psi.$ This can be absorbed
by the coordinate redefinition $du\rightarrow du-\frac{q}{m}d\psi.$
For the single string, $u$ is periodic, hence we need that 
\begin{equation}
\frac{4\pi q}{lm}\in\mathbb{Z},
\end{equation}
where $l$ is the length of the circle, for this to be well-defined.
The limit
\begin{equation}
\lim_{r\rightarrow\infty}H=\sqrt{\Omega_{\alpha\beta}\mu_{\infty}^{\alpha}\mu_{\infty}^{\alpha}}=1
\end{equation}
then implies that
\begin{equation}
\Omega_{\alpha\beta}\mu_{\infty}^{\alpha}\mu_{\infty}^{\beta}=1.\label{condition radius}
\end{equation}
Lastly, the limit of the one-form $\omega\rightarrow0$ implies that
\begin{equation}
\lim_{r\rightarrow0}\omega_{0}=j_{\infty}+\frac{1}{m}\Omega_{\alpha\beta}\mu_{\infty}^{\alpha}p^{\beta}-\frac{1}{2}\frac{q}{m}=0
\end{equation}
such that 
\begin{equation}
j_{\infty}=\frac{1}{2}\frac{q}{m}-\frac{1}{m}\Omega_{\alpha\beta}\mu_{\infty}^{\alpha}p^{\beta}.
\end{equation}
Hence, to get the correct asymptotics, we need 
\begin{equation}
\Gamma_{\infty}=\left(\mu_{\infty}^{\alpha},0,0,0,-1,\frac{1}{2}\frac{q}{m}-\frac{1}{m}\Omega_{\alpha\beta}\mu_{\infty}^{\alpha}p^{\beta}\right)
\end{equation}
subject to (\ref{condition radius}). With these values for $\Gamma_{\infty},$
the Bubble equations are automatically satisfied.

\section{Reduction of the flow equation to five dimensions \label{Sec. reduction flow equation to five dim}}

We  show that when compactifying along the $u-$circle (as done in
section \ref{subsec:Reduction-to-five-dimensions}), the flow equation
(\ref{Flow equation u-indep}) reduces to the flow equation derived
in \cite{Kraus:2005gh}, which in our conventions is given by\footnote{In their conventions $X_{I}\equiv\frac{1}{2}C_{IJK}X^{J}X^{K}$ and
$\alpha\cdot\beta=\alpha_{mn}\beta^{mn}.$}
\begin{equation}
\tilde{\nabla}^{l}\left(f^{-1}\mathcal{G}_{IJ}X^{I}E_{l}^{J}\right)=f^{-1}\mathcal{G}_{IJ}\partial_{l}X^{I}\partial^{l}X^{J}-\frac{1}{4}C_{IJK}X^{I}\Theta^{J}\cdot\Theta^{K},
\end{equation}
where
\begin{equation}
\mathcal{G}_{IJ}=\left[-\frac{1}{2}\partial_{X^{I}}\partial_{X^{J}}\log\mathcal{N}\right]|_{\mathcal{N}=1}=\frac{9}{2}X_{I}X_{J}-\frac{1}{2}C_{IJK}X^{K}
\end{equation}
and $E_{l}^{I}=F_{lv}^{I}=f^{-1}\partial_{l}\left(fX^{I}\right)$.
Some useful identities that follow from $X_{I}X^{I}=1$ are:
\begin{eqnarray}
\mathcal{G}_{IJ}X^{J} & = & \frac{3}{2}X_{I},\nonumber \\
\partial_{l}X_{I} & = & -\frac{2}{3}\mathcal{G}_{IJ}\partial_{l}X^{J}.\label{eq:identities 5dsugra}
\end{eqnarray}

From the cubic potential (\ref{cubic potential in 6d data}) one finds

\begin{eqnarray}
\mathcal{G}_{\alpha\beta} & = & r^{-4/3}g_{\alpha\beta},\nonumber \\
\mathcal{G}_{0\beta} & = & 0,\label{metric scalar space 5d}\\
\mathcal{G}_{00} & = & \frac{1}{2}r^{8/3}.\nonumber 
\end{eqnarray}
Let's now reduce the terms in the flow equation (\ref{Flow equation u-indep})
one by one. We start with the left-hand side:
\begin{equation}
\frac{1}{6}\tilde{\nabla}_{l}\left(\Omega_{\alpha\beta}j^{\alpha}\hat{G}_{ijk}^{\beta}\right)\epsilon^{lijk}.
\end{equation}
Note that the base space in the five- and six-dimensional space has
the same metric $ds_{4}^{2},$ so the covariant derivative does not
change. The vierbein $\hat{e}^{i}$ is related to a vierbein $e^{i}$
of the 5d spatial part via $\hat{e}^{i}=r^{-1/3}e^{i}$ and $e^{i}=f^{-1/2}\tilde{e}^{i}.$
Applying the self-duality condition (\ref{Self-duality condition})
to the ansatz for the three-forms (\ref{ansatz three-forms}) relates
$G^{\alpha}$ to the two-forms $F^{\alpha}.$ In particular when we
express
\begin{equation}
G^{\alpha}=\frac{1}{2}G_{ij}^{\alpha}\tilde{e}^{i}\wedge\tilde{e}^{j}\wedge\left(dv+\omega\right)+\frac{1}{6}G_{ijk}^{\alpha}\tilde{e}^{i}\wedge\tilde{e}^{j}\wedge\tilde{e}^{k},
\end{equation}
we find that
\begin{equation}
G_{ijk}^{\alpha}=-f^{-1}r^{-4/3}g^{\alpha\beta}\Omega_{\beta\gamma}E_{l}^{\gamma}\epsilon_{\ ijk}^{l}.
\end{equation}
Hence 
\begin{equation}
\hat{G}_{ijk}^{\alpha}=G_{ijk}^{\alpha}=-f^{-1}r^{-4/3}\left(2j^{\alpha}j_{\gamma}-\delta_{\gamma}^{\alpha}\right)E_{l}^{\gamma}\epsilon_{\ ijk}^{l}.
\end{equation}
We then derive that
\begin{eqnarray}
\frac{1}{6}\tilde{\nabla}_{l}\left(\Omega_{\alpha\beta}j^{\alpha}\hat{G}_{ijk}^{\beta}\right)\epsilon^{lijk} & = & -\tilde{\nabla}^{l}\left(f^{-1}r^{-4/3}\Omega_{\alpha\beta}j^{\alpha}E_{l}^{\beta}\right).
\end{eqnarray}
Inserting the ansatz for the scalars (\ref{ansatz scalars}) and applying
the product rule gives
\begin{eqnarray}
\frac{1}{6}\tilde{\nabla}_{l}\left(\Omega_{\alpha\beta}j^{\alpha}\hat{G}_{ijk}^{\beta}\right)\epsilon^{lijk} & = & -r^{-2/3}\tilde{\nabla}^{l}\left(f^{-1}r^{-4/3}\Omega_{\alpha\beta}X^{\alpha}E_{l}^{\beta}\right)-f^{-1}r^{-4/3}\Omega_{\alpha\beta}X^{\alpha}E_{l}^{\beta}\partial^{l}\left(r^{-2/3}\right).\nonumber \\
\label{first term flow eq v1}
\end{eqnarray}
Using (\ref{metric scalar space 5d}) we then calculate that
\begin{eqnarray}
\tilde{\nabla}^{l}\left(f^{-1}\mathcal{G}_{IJ}X^{I}E_{l}^{J}\right) & = & \frac{1}{2}\tilde{\nabla}^{l}\left(f^{-1}r^{4/3}E_{l}^{0}\right)+\tilde{\nabla}^{l}\left(f^{-1}r^{-4/3}\Omega_{\alpha\beta}X^{\alpha}E_{l}^{\beta}\right)\nonumber \\
\label{first term flow eq v1 part 1}
\end{eqnarray}
and combining (\ref{metric scalar space 5d}) with the definition
of $E_{l}^{\beta}$ and $\partial_{l}r=-\frac{3}{4}r^{7/3}\partial_{l}X^{0},$
the second term of the right-hand side in (\ref{first term flow eq v1})
can be calculated:
\begin{eqnarray}
f^{-1}r^{-4/3}\Omega_{\alpha\beta}X^{\alpha}E_{l}^{\beta}\partial^{l}\left(r^{-2/3}\right) & = & -\frac{1}{2}r^{-2/3}f^{-1}\mathcal{G}_{00}\partial_{l}\left(X^{0}\right)\partial^{l}\left(X^{0}\right)-\frac{2}{3}f^{-2}r^{-5/3}\partial_{l}\left(f\right)\partial^{l}\left(r\right).\nonumber \\
\label{second term flow eq v1 part 2}
\end{eqnarray}
Substitution of (\ref{first term flow eq v1 part 1}) and (\ref{second term flow eq v1 part 2})
in (\ref{first term flow eq v1}) yields
\begin{eqnarray}
\frac{1}{6}\tilde{\nabla}_{l}\left(\Omega_{\alpha\beta}j^{\alpha}\hat{G}_{ijk}^{\beta}\right)\epsilon^{lijk} & = & -r^{-2/3}\tilde{\nabla}^{l}\left(f^{-1}\mathcal{G}_{IJ}X^{I}E_{l}^{J}\right)+\frac{1}{2}r^{-2/3}\tilde{\nabla}^{l}\left(f^{-1}r^{4/3}E_{l}^{0}\right)\nonumber \\
 &  & +\frac{1}{2}r^{-2/3}f^{-1}\mathcal{G}_{00}\partial_{l}\left(X^{0}\right)\partial^{l}\left(X^{0}\right)+\frac{2}{3}f^{-2}r^{-5/3}\partial_{l}\left(f\right)\partial^{l}\left(r\right).\nonumber \\
\label{eq:first part flow result}
\end{eqnarray}

Let us then reduce the first term of the right-hand side of the flow
equation (\ref{Flow equation u-indep}). Inserting the ansatz for
the scalars (\ref{ansatz scalars}), applying the product rule and
using (\ref{metric scalar space 5d}) yields
\begin{eqnarray}
Hg_{\alpha\beta}\partial_{l}\left(j^{\alpha}\right)\partial^{l}\left(j^{\beta}\right) & = & H\mathcal{G}_{\alpha\beta}\partial_{l}\left(X^{\alpha}\right)\partial^{l}\left(X^{\beta}\right)+2Hr^{2/3}\mathcal{G}_{\alpha\beta}X^{\beta}\partial_{l}\left(X^{\alpha}\right)\partial^{l}\left(r^{-2/3}\right)\nonumber \\
 &  & +\frac{4}{9}Hr^{-2}\partial_{l}\left(r\right)\partial^{l}\left(r\right).\label{second part flow result v1}
\end{eqnarray}
Using that $H=r^{-2/3}f^{-1},$ 
\begin{equation}
\mathcal{G}_{\alpha\beta}X^{\beta}\partial_{l}\left(X^{\alpha}\right)=\frac{2}{3}r^{-1}\partial_{l}r
\end{equation}
and
\begin{equation}
H\mathcal{G}_{00}\partial_{l}\left(X^{0}\right)\partial^{l}\left(X^{0}\right)=\frac{8}{9}Hr^{-2}\partial_{l}\left(r\right)\partial^{l}\left(r\right),
\end{equation}
we find that 
\begin{eqnarray}
Hg_{\alpha\beta}\partial_{l}\left(j^{\alpha}\right)\partial^{l}\left(j^{\beta}\right) & = & r^{-2/3}f^{-1}\mathcal{G}_{\alpha\beta}\partial_{l}\left(X^{\alpha}\right)\partial^{l}\left(X^{\beta}\right)-\frac{1}{2}r^{-2/3}f^{-1}\mathcal{G}_{00}\partial_{l}\left(X^{0}\right)\partial^{l}\left(X^{0}\right).\nonumber \\
\label{second part flow result}
\end{eqnarray}

Lastly, we reduce the second term of the right-hand side of the flow
equation (\ref{Flow equation u-indep}). Using (\ref{cubic potential in 6d data})
and (\ref{Self-dual tensors 4d}) we can expand
\begin{eqnarray}
-\frac{1}{4}C_{IJK}X^{I}\Theta^{J}\cdot\Theta^{K} & = & -\frac{1}{2}\Omega_{\alpha\beta}r^{-4/3}\mathcal{G}^{+\alpha}\cdot\mathcal{G}^{+\beta}+r^{2/3}j_{\alpha}\tilde{d}\beta\cdot\mathcal{G}^{+\alpha}.\label{eq:theta part}
\end{eqnarray}
For the first term at the right-hand side of (\ref{eq:theta part})
we use the reduced Einstein equation (\ref{eq: reduced Einstein equation}):
\begin{equation}
\frac{1}{3}\Omega_{\alpha\beta}\mathcal{G}^{+\alpha}\cdot\mathcal{G}^{+\beta}=\tilde{\nabla}^{2}\left(f^{-1}X_{0}\right)=-\frac{2}{3}\tilde{\nabla}^{l}\left(f^{-1}\mathcal{G}_{00}E_{l}^{0}\right),
\end{equation}
where the second equality follows using (\ref{eq:identities 5dsugra})
and the definition of $E_{l}^{0}$. Inserting $\mathcal{G}_{00}$
and using again the definition of $E_{l}^{0}$ we find that
\begin{eqnarray}
-\frac{1}{2}\Omega_{\alpha\beta}r^{-4/3}\mathcal{G}^{+\alpha}\cdot\mathcal{G}^{+\beta} & = & r^{-4/3}\tilde{\nabla}^{l}\left(f^{-1}\mathcal{G}_{00}E_{l}^{0}\right)\nonumber \\
 & = & \frac{1}{2}\tilde{\nabla}^{l}\left(f^{-1}r^{4/3}E_{l}^{0}\right)+\frac{1}{2}f^{-1}\partial_{l}\left(X^{0}\right)\partial^{l}\left(r^{4/3}\right)+\frac{1}{2}f^{-2}X^{0}\partial_{l}\left(f\right)\partial^{l}\left(r^{4/3}\right)\nonumber \\
 & = & \frac{1}{2}\tilde{\nabla}^{l}\left(f^{-1}r^{4/3}E_{l}^{0}\right)-f^{-1}\mathcal{G}_{00}\partial_{l}\left(X^{0}\right)\partial_{l}\left(X^{0}\right)+\frac{2}{3}r^{1/3}f^{-2}X^{0}\partial_{l}\left(f\right)\partial^{l}\left(r\right).\nonumber \\
\label{first part last term flow}
\end{eqnarray}
Substitution of (\ref{first part last term flow}) in (\ref{eq:theta part})
yields
\begin{eqnarray}
-j_{\alpha}\tilde{d}\beta\cdot\mathcal{G}^{+\alpha} & = & -r^{-2/3}\Biggl[-\frac{1}{4}C_{IJK}X^{I}\Theta^{J}\cdot\Theta^{K}-\frac{1}{2}\tilde{\nabla}^{l}\left(f^{-1}r^{4/3}E_{l}^{0}\right)+f^{-1}\mathcal{G}_{00}\partial_{l}\left(X^{0}\right)\partial^{l}\left(X^{0}\right)\nonumber \\
 &  & -\frac{2}{3}f^{-2}r^{-1}\partial_{l}\left(f\right)\partial^{l}\left(r\right)\Biggl].\label{eq:last part flow result}
\end{eqnarray}
Substitution of (\ref{eq:first part flow result}), (\ref{second part flow result})
and (\ref{eq:last part flow result}) in (\ref{Flow equation u-indep})
yields
\begin{eqnarray}
\tilde{\nabla}^{l}\left(f^{-1}\mathcal{G}_{IJ}X^{I}E_{l}^{J}\right) & = & f^{-1}\mathcal{G}_{IJ}\partial_{l}\left(X^{I}\right)\partial^{l}\left(X^{J}\right)-\frac{1}{4}C_{IJK}X^{I}\Theta^{J}\cdot\Theta^{K}.
\end{eqnarray}
This is the five-dimensional flow equation.

\bibliographystyle{elsarticle-num}
\bibliography{references}

\end{document}